\documentclass[review,3p,12pt]{elsarticle}

\usepackage{amssymb}
\usepackage{amsmath}
\usepackage{booktabs} 
\usepackage[hidelinks,linktoc=all]{hyperref} 
\usepackage{xcolor}
\newcommand{\rev}[1]{\textcolor{black}{#1}} 

\journal{Journal of Colloid and Interface Science}

\begin{document}

\begin{frontmatter}

\title{Unmasking the internal structure of casein micelles through enzymatic hydrolysis: A SAXS study}


\author[LYON]{Julien Bauland\corref{cor1}}
\ead{julien.bauland@univ-lyon1.fr}

\author[RENNES]{Ghazi Ben Messaoud}

\author[SACLAY]{François Boué}

\author[RENNES]{Pascaline Hamon}

\author[RENNES]{Florence Rousseau}

\author[LYON1]{Thomas Gibaud}

\author[RENNES]{Thomas Croguennec\corref{cor1}}
\ead{thomas.croguennec@institut-agro.fr}

\affiliation[LYON]{organization={Universite Claude Bernard Lyon 1, CNRS, Institut Lumière Matière, UMR5306},
            addressline={F-69100}, 
            city={Villeurbanne},
            country={France}}

\affiliation[RENNES]{organization={INRAE, Institut Agro, STLO},
            addressline={65 Rue de Saint Brieuc, F-35042}, 
            city={Rennes},
            country={France}}

\affiliation[SACLAY]{organization={Laboratoire Léon Brillouin, UMR12 CEA-CNRS, Université Paris-Saclay, CEA Saclay},
            addressline={F-91191}, 
            city={Gif sur Yvette},
            country={France}}

\affiliation [LYON1]{Univ Lyon, Ens de Lyon, CNRS, Laboratoire de Physique, 69342 Lyon, France}
            
\cortext[cor1]{Corresponding authors} 





\end{frontmatter}


\section{Introduction} 

Constituting the main protein fraction of bovine milk, \emph{caseins} are phosphorylated proteins with an average molecular weight of $20~\rm{kDa}$, present in the milk of all mammals~\citep{holt2013}. Despite their lack of well-defined tertiary structure, caseins are naturally assembled into large quaternary structures of approximately $100{,}000~\rm{kDa}$, commonly referred to as \emph{casein micelles}. Casein micelles are roughly spherical colloids with a highly hydrated structure (2--4 g of water per g of protein) and can be viewed as protein microgels with a typical diameter of $100~\rm{nm}$ \citep{fox2008casein,HORNE2020213}. A key feature of these assemblies is the presence of nanometric \rev{clusters of amorphous} calcium phosphate, referred to as ``calcium phosphate nanoclusters'', which are stabilized by the abundant phosphoseryl residues present in the primary sequence of caseins \citep{holt1986,holt2004equilibrium}.

The composition and structure of casein micelles are closely linked to their two main biological functions~\citep{holt2013}. Firstly, the sequestration of calcium phosphate nanoclusters enables the delivery of large amounts of calcium and phosphate to the neonate, while preventing pathological calcification of mammary glands during milk secretion. Secondly, the presence of a specific casein fraction, namely $\kappa$-casein, ensures the colloidal stability of casein micelles in milk. In the stomach of the neonate, the secretion of a specific proteolytic enzyme, \emph{chymosin}, which selectively cleaves $\kappa$-casein, leads to the loss of the repulsion between casein micelle, hence of their colloidal stability, and the subsequent gelation of the bolus. The overall effect is enhancing protein digestion.

Beyond its biological role, the ability of milk, i.e., a dispersion of casein micelles, to gel upon \emph{chymosin} (or ``rennet'') addition has been exploited since the end of the prehistoric times to increase milk preservation through cheese manufacture~\citep{salque2013}. Today, enzymatic gelation and the subsequent cutting of the gel to expel whey, remain critical steps in cheese production, as they determine both processing yield and the final properties of the cheese~\citep{bauland2024chapter}. Enzymatic milk gelation is a multi-step process involving (i) enzymatic hydrolysis of $\kappa$-casein, (ii) aggregation of destabilized casein micelles, and (iii) gel formation and aging~\citep{fox2016,horne2017}. As reflected by the diversity of cheese varieties, enzymatic milk gelation can occur under a wide range of physicochemical conditions, including pH, temperature, and ionic composition. As dynamic colloidal objects, casein micelles, the fundamental building blocks of dairy matrices, readily adapt their structure and composition to these conditions \citep{gaucheron2005minerals}. This intrinsic versatility, coupled with varying processing conditions, makes the control and standardization of dairy products particularly challenging.

More generally, the physico-chemistry of casein micelles governs the processing of a broad range of food products, including cheeses, yogurts, milk concentrates, and milk powders~\citep{lucey2003}. Understanding how the microscopic properties of these building blocks control the macroscopic properties of food matrices therefore requires a detailed knowledge of casein micelle structure and of its evolution in response to environmental changes~\citep{horne2020}.

Driven by this need, numerous studies have investigated the structure of casein micelles, primarily using electron microscopy~\citep{marchin2007effects,trejo2011cryo} and small-angle X-ray or neutron scattering~\citep{holt2003substructure,DeKruif2003,de2012casein,de2014structure,ingham2016revisiting,day2017probing,pedersen2022model,smith2020casein,moller2025structural}. Major experimental challenges arise from the small size and the dynamic nature of the colloids, which can be altered by the fixation procedures often required for electron microscopy observations. Small-angle scattering patterns show a great suitability for investigations at this scale, especially \emph{in situ}. However, additional difficulties come from the non-uniqueness of interpretation. As a consequence, in spite of the general acceptation recalled above (i) $\kappa$-casein \rev{ensures the stability of native casein micelle suspensions} and (ii) calcium phosphate nanoclusters are stabilized by clusters of phosphoseryl residues, the precise spatial organization of the different casein fractions and mineral components within the micelle remains a matter of debate.

To gain further insight into casein micelle structure, one strategy has been to probe casein micelle structural modifications in response to controlled environmental perturbations, such as changes in pH \citep{marchin2007effects, lazzaro2020tailoring, li2022situ, takagi2024saxs} or osmotic pressure \citep{bouchoux2010squeeze}. In this work, we adopt a similar approach by investigating the structure of casein micelles during the enzymatic gelation of \rev{a reconstituted skimmed milk powder} using small-angle X-ray scattering (SAXS). By combining two sample-to-detector distances, we perform time-resolved measurements to track structural changes throughout the successive stages of enzymatic gelation over length scales ranging from $3~\rm{nm}$ to $3$~\textmu m. Complementary techniques, namely peptide titration, dynamic light scattering, and rheometry, are used to characterize the kinetics of enzymatic gelation, providing temporal references to support the analysis of the SAXS data.

Using a model-free analysis of the SAXS spectra, we show that, prior to any aggregation, $\kappa$-casein hydrolysis leads to a decrease in the scattering intensity at intermediate length scales, concomitant with an increase in the intensity of the casein micelle form factor. We interpret this result as a hydrolysis-driven relaxation of the porous internal structure of casein micelles, providing, for the first time, direct evidence of the central role of $\kappa$-casein in defining the intermediate structural level of the colloid. This view is further supported by a comparison of our data with the two existing structural models for the casein micelle substructure. Furthermore, we show that the progressive reduction of the scattering at intermediate length scales during the hydrolysis and aggregation of casein micelles reveals a structural peak at small length scales, typically observed by contrast-matching techniques in neutron scattering. These results show the potential of enzymatic gelation to provide complementary mean of studying the fine structural organization of casein micelles.


\section{Materials and Methods}  

\subsection{Sample Preparation}

Skim milk was reconstituted at 100~g.L$^{-1}$ in ultrapure water from a low-heat skim milk powder (Armor Protéines, France), resulting in a casein micelle volume fraction of $\approx 10\%$~\citep{bauland2024}. The "low-heat" specification ensures minimal heat treatment during drying, thereby limiting soluble protein denaturation and maintaining the reconstituted powder as close as possible to its native state. The skim milk powder composition is summarized in Table~\ref{tab:composition}. After powder dispersion in water, the casein concentration, determined by the Kjeldahl method, was 25.8~g.L$^{-1}$.

\begin{table}[h]
\centering
\caption{Composition of the skim milk powder.}
\label{tab:composition}
\begin{tabular}{lc|lc}
\hline
Component & Content & Component & Content (g.kg$^{-1}$) \\ \hline
Total Protein & 32.2\% & Potassium & 14.93 \\
Calcium & 10.42 g.kg$^{-1}$ & Sodium & 4.56 \\
Magnesium & 1.11 g.kg$^{-1}$ & Citrate & 16.6 \\
Phosphate & 17.5 g.kg$^{-1}$ & Chloride & 9.5 \\ \hline
\end{tabular}
\end{table}

The milk was supplemented with 0.2~g.L$^{-1}$ sodium azide as a biopreservative, adjusted to pH 6.6 using 1~M HCl, and stirred overnight at room temperature.

To initiate enzymatic gelation, milk samples were warmed to $38^{\circ}$C, before adding a fermentation-produced camel \emph{chymosin} (CHY-MAX Supreme, Chr-Hansen, Denmark; 1000\allowbreak~IMCU.mL$^{-1}$). This enzyme cleaves $\kappa$-casein at the Phe105-Met106 peptide bond, thereby releasing the C-terminal part of $\kappa$-casein (CMP) in the whey. It provides enhanced affinity for $\kappa$-casein compared to bovine \emph{chymosin}, effectively minimizing non-specific proteolytic activity. The coagulant was added at a final concentration of 1~IMCU per gram of casein. Samples were gently mixed and immediately loaded into the various measurement devices, which were maintained at $38^{\circ}$C throughout the gelation process.

\subsection{Monitoring casein micelles hydrolysis, aggregation and gelation}

The kinetics of $\kappa$-casein hydrolysis, micelle aggregation, and gel formation were monitored via RP-HPLC, dynamic light scattering (DLS), and small-oscillatory rheometry, respectively. To monitor the kinetics following enzyme addition at 38$^{\circ}$C, specific quenching procedures were employed to ensure accurate temporal snapshots of the structural state.

For $\kappa$-casein hydrolysis, 2~mL of milk sample was added to 4~mL of 3\% trichloroacetic acid to stop the reaction at different defined times,~\citep{lauzin2019effect}. The mixture was kept at rest at room temperature for 10~min to enable protein precipitation and then centrifuged at 7000~g for 10~min. The supernatant was recovered for RP-HPLC analysis to determine CMP release. The CMP release is represented by $\chi_{enz}$ which corresponds to the ratio of the area under the RP-HPLC chromatographic peak of CMP at time t to that at infinite time.

 The micelle aggregation was quenched by sampling 200~$\mu$L of chymosin-treated milk and diluting it in 0.8~mL of cooled water. The mixture was kept at 4$^{\circ}$C until particle size measurement by dynamic light scattering.
 
For rheometry and USAXS/SAXS characterization, the samples were transferred to the required measurement devices (rheometer and flow-through cell) rapidly after \emph{chymosin} addition. The delay between the time of \emph{chymosin} addition and the starting time of data acquisition was measured and added to the total acquisition time.

\subsubsection{Reverse-Phase HPLC}
The caseinomacropeptide (CMP, N-term peptide fragment [106-169] of $\kappa$-casein) was quantified on a PLRP-S column (8~$\mu$m, 150 $\times$ 2~mm, Agilent Technologies, Santa Clara, CA) connected to a Waters chromatography system (Milford, USA). The system consisted of a Waters 2695 Separation Module, a Waters 2487 Dual Wavelength Absorbance Detector, and Empower chromatography application software. The column was eluted at 40$^{\circ}$C at a constant flow rate of 0.2~mL.min$^{-1}$ using a nonlinear gradient of acetonitrile with solvent A (99.9\% water, 0.1\% trifluoroacetic acid) and solvent B (99.9\% acetonitrile, 0.1\% trifluoroacetic acid). After a plateau of 2~min with 16\% of buffer B, the proportion of buffer B increased from 16\% to 32\% in 3~min, then to 48\% in 15~min, and finally to 80\% in 1~min. CMP was detected at $\lambda = 214$~nm.

\subsubsection{Dynamic Light Scattering}
Particle size was measured using a Zetasizer NanoZS apparatus (Malvern Instruments, Worcestershire, UK) equipped with a 4~mW He/Ne laser operating at a wavelength of 633~nm and an automatic attenuator. Prior to analysis, samples were diluted 100-fold in water to minimize multiple scattering effects and equilibrated at 20$^{\circ}$C. Measurements were performed in disposable polystyrene cuvettes (10 $\times$ 10~mm$^{2}$), and each sample was analyzed in triplicate. The translational diffusion coefficient was obtained from intensity fluctuations in backscattering mode at a detection angle of 173$^{\circ}$. The hydrodynamic diameter ($Z$-average) was calculated using the Stokes–Einstein equation, assuming the viscosity of water at 20$^{\circ}$C.

\subsubsection{Small Amplitude Oscillatory Shear Rheometry}
Rheological experiments were performed using a stress-controlled rheometer (MCR 301, Anton Paar, Austria) equipped with a concentric cylinder geometry (CC17; cup diameter = 18.08~mm, bob diameter = 16.66~mm). The gelation kinetics were monitored by following the temporal evolution of the storage ($G^{\prime}$) and loss ($G^{\prime\prime}$) moduli over 90~min at 38$^{\circ}$C. Measurements were carried out at a constant angular frequency $\omega = 0.628~\text{rad.s}^{-1}$ ($f = 0.1$~Hz) and a fixed strain amplitude $\gamma_0 = 0.01$. The minimum measurable viscoelastic modulus ($G_{\min}$) was estimated from the minimum detectable torque ($T_{\min} = 0.01~\mu\text{N.m}$) according to $G_{\min} = (k_{\sigma} \, T_{\min}) / \gamma_0$, where $k_{\sigma} = 1 / (2\pi R_i^2 L)$ is the geometric stress conversion factor. For this geometry ($R_i = 8.33$~mm, $L \approx 50$~mm), $G_{\min}$ was estimated to be $\approx 4.59 \times 10^{-2}$~Pa.

\subsection{Small-Angle X-ray Scattering (SAXS)}

Small-angle X-ray scattering experiments were conducted at the ID02 beamline of the European Synchrotron Radiation Facility (ESRF, Grenoble, France)~\citep{narayanan2022}. Two-dimensional scattering patterns were acquired using a flow-through-cell setup maintained at $38^{\circ}$C. Enzymatic gelation was monitored for 90~min with an acquisition frequency of one frame every 20~s,\rev{with an exposure time
of 100~ms per frame,} using an X-ray energy of 12.23~keV. To cover a wide scattering wave-vector range, $2 \times 10^{-3}\leq q \leq 3~\text{nm}^{-1}$, measurements were performed separately at two sample-to-detector distances of 2.5 and 30.68~m. The scattering wave vector is defined as $q = (4\pi \sin\theta) / \lambda$, where $\lambda$ is the X-ray wavelength and $2\theta$ is the scattering angle.

One-dimensional scattering curves, $I(q)$, were obtained by azimuthal averaging of the 2D isotropic patterns after masking artifacts and the beam-stop shadow. Background subtraction was performed independently for each detector distance using the scattering spectrum of the supernatant obtained after ultracentrifugation of the gel, corresponding to the aqueous phase devoid of casein micelles (see Appendix for further details). 

The background-corrected SAXS and USAXS profiles acquired at equal value of elapsed time after triggering the enzymatic reaction were subsequently merged using their q overlap region. Prior such stitching, the SAXS profiles were rescaled to match the corresponding USAXS profiles. The scaling factors remained close to unity throughout the experiment, ensuring consistent reconstruction of the scattering curves over the full accessible q-range. 

\section{Results and Discussion} 

\subsection{Evolution of SAXS Spectra During Enzymatic Gelation}
\label{sec:gel}

\begin{figure*}[t!]
    \includegraphics[scale=0.4, clip=true, trim=0mm 0mm 0mm 0mm]{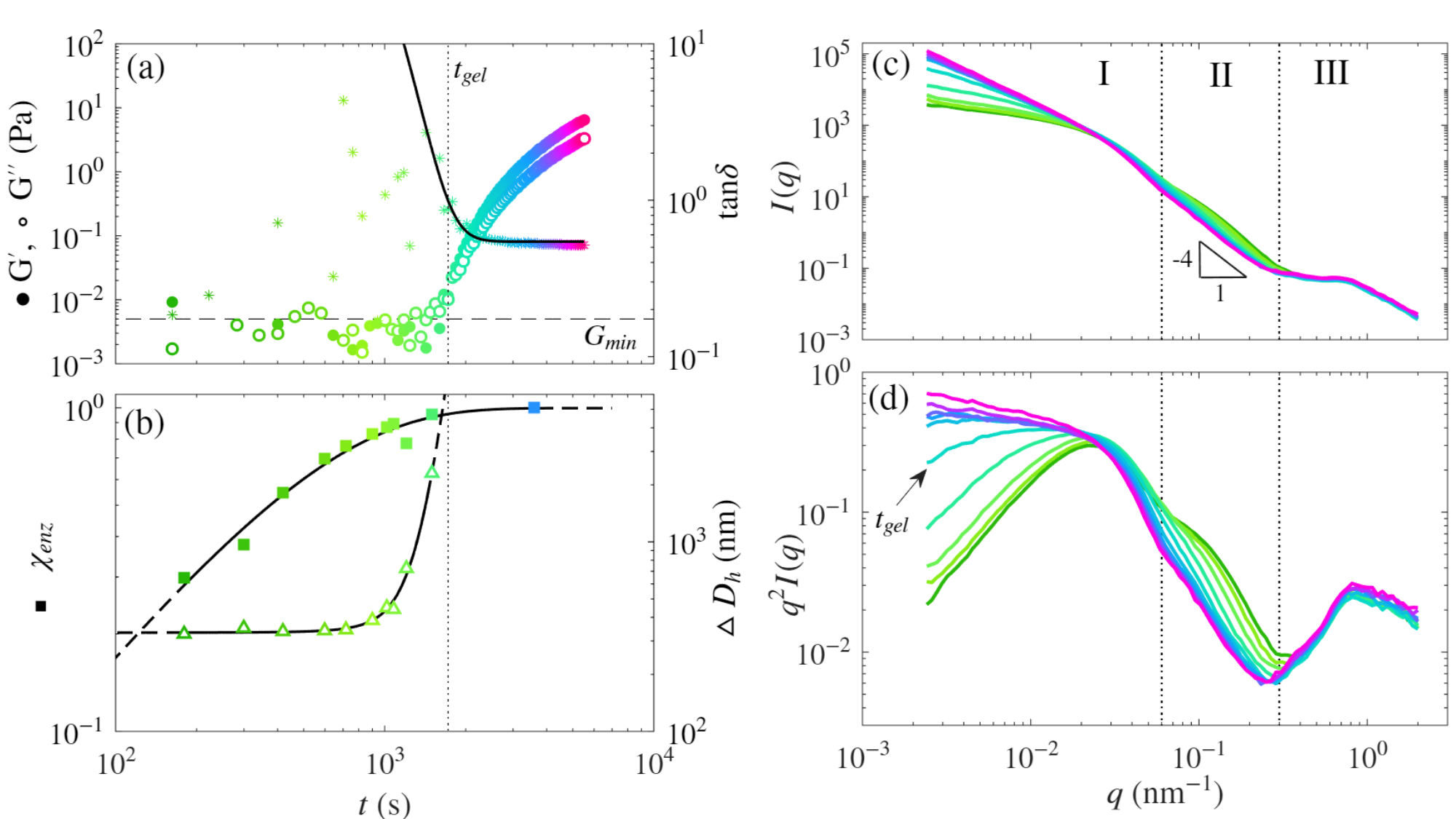}
    \centering
    \caption{(a) Temporal evolution of the viscoelastic moduli (left axis), $G^{\prime}$ and $G^{\prime\prime}$, and of $\tan\delta$ (right axis) following enzyme addition. The black curve corresponds to the best exponential fit, $\tan\delta(t)=\exp[(t_{\mathrm{gel}}-t)/\tau]$. The vertical dashed line indicates the effective gel time, defined by $G^{\prime}(t_{\mathrm{gel}})=G^{\prime\prime}(t_{\mathrm{gel}})$. The horizontal dashed line indicates the minimum measurable shear modulus. (b) Extent of the enzymatic reaction $\chi_{\mathrm{enz}}$, assessed from titration of caseinomacropeptide release (left axis), and hydrodynamic diameter $D_h$ measured by dynamic light scattering, as a function of time after enzyme addition. Black lines show the best exponential fits: $\chi_{\mathrm{enz}}(t)=1-\exp(-t/\tau)$ and $D_h(t)=D_{\infty}[1-\exp(-t/\tau)]+D_0$. (c) Evolution of the azimuthally averaged scattered intensity $I(q)$ as a function of the scattering wave vector $q$. The color codes for the time after enzyme addition, with the same tones as in (a,b). Vertical dashed lines delimit three structural regimes: low-$q$ (I), intermediate-$q$ (II), and high-$q$ (III). (d) Corresponding Kratky representation, $q^{2}I(q)$. The arrow indicates the gel time determined from rheology.
    }
    \label{fig:gel}
\end{figure*}

Fig.~\ref{fig:gel}(a)--(b) illustrates the three successive steps of enzymatic gelation, namely (i) hydrolysis, (ii) aggregation, and (iii) gelation, as assessed by titration of the enzymatic reaction, DLS, and rheometry, respectively. In Fig.~\ref{fig:gel}(b), the extent of enzymatic reaction $\chi_{enz}$ (ratio of the amount of CMP released at time $t$ to the amount released at infinite time) is well described by a first-order kinetics,
$\chi_{enz} = 1 - e^{-t/\tau}$, with a characteristic time $\tau = 540~\rm{s}$, corresponding to approximately $63~\%$ of the reaction completion. When $\chi_{enz} \approx 0.85$, reached at $t \simeq 1000~\rm{s}$ after enzyme addition, DLS shows that the hydrodynamic diameter $D_h$ starts to increase, reflecting the loss of colloidal stability of casein micelles and the onset of aggregation. This result is consistent with previous studies reporting destabilization of casein micelles at pH 6.6 once about $80-90~\%$ of the $\kappa$-casein fraction has been hydrolyzed~\citep{horne2017}. The exponential increase of $D_h$ is consistent with a reaction-limited cluster aggregation mechanism~\citep{zaccone2010,Begam2021}, ultimately leads to the formation of a percolated network, corresponding to the true gel point, where the attractive dispersion becomes solid-like.

The sol--gel transition is evidenced in Fig.~\ref{fig:gel}(a), which shows the temporal evolution of the elastic modulus $G^{\prime}$ and the viscous modulus $G^{\prime\prime}$ measured at fixed frequency. At $t \simeq 1750~\rm{s}$, $G^{\prime}$ exceeds $G^{\prime\prime}$, indicating the onset of solid-like behavior and providing an estimate of the gel time, hereafter denoted $t_{gel}$. At $t_{gel}$, more than 95~\% of the $\kappa$-casein fraction has been hydrolyzed [Fig.~\ref{fig:gel}(b)], in agreement with previous reports~\citep{sandra2012effect}. For $t > t_{gel}$, the continued increase of both $G^{\prime}$ and $G^{\prime\prime}$ has been attributed to endogenous structural aging and densification of the fractal network formed at the gel point~\citep{Mellema2002,bauland2024}.

Fig.~\ref{fig:gel}(c) displays the temporal evolution of the one-dimensional scattering intensity $I(q)$ as a function of the scattering wave vector $q$, measured for the casein micelle dispersion during enzymatic gelation [same color code as in Fig.~\ref{fig:gel}(a)--(b)]. The dark green curve, associated with the stable dispersion of native casein micelles, and the pink curve, associated with the aged gel, show typical SAXS profiles of casein micelle dispersions~\citep{bouchoux2010squeeze, ingham2016revisiting,  smith2020casein, pedersen2022model} and casein gels~\citep{li2018rheological,bauland2024}, respectively. The scattered intensity probes structural features at length scales $\xi = 2\pi/q$. To guide the analysis, the spectra are divided into three $q$-regimes, denoted I (low-$q$, $q < 6 \times 10^{-2}~\rm{nm^{-1}}$), II (intermediate-$q$, $6 \times 10^{-2} < q < 3 \times 10^{-1}~\rm{nm^{-1}}$), and III (high-$q$, $q > 3 \times 10^{-1}~\rm{nm^{-1}}$), as indicated by vertical dashed lines.

First, we focus on the low-$q$ regime, corresponding to large length scales. The dark green curve, associated with the stable dispersion of native casein micelles, exhibits a break in slope at $q \simeq 3 \times 10^{-2}~\rm{nm^{-1}}$, commonly attributed to the spherical form factor of casein micelles~\citep{ingham2016revisiting}. This feature is more clearly evidenced in the Kratky representation, $q^{2}I(q)$ vs. $q$, where breaks in slope appear as maxima or minima. As the enzymatic reaction proceeds, the increase in scattered intensity at $q < 3 \times 10^{-2}~\rm{nm^{-1}}$ reflects casein micelles aggregation and network formation. Interestingly, before the onset of casein micelle aggregation a first gentle intensity increase is noticeable but does not change the bell shape. Then, a flat curve is reached slightly above $t_{gel}$, indicating the ongoing percolation of the casein micelle clusters. At longer time, $I(q)$ vs. $q$ approaches asymptotically a straight line of negative slope. For a fractal organization characterized by a fractal dimension $d_f$, the scattering intensity follows a power-law scaling $I(q) \propto q^{-d_f}$~\citep{Schmidt1991}. At the gel time $t_{gel}$, as determined from rheology (light blue curve indicated by an arrow), the low-q part of the SAXS spectrum shows a power-law regime with a power exponent $d_f \simeq 2$--$2.1$, thus appearing as a horizontal line in Kratky representation. Beyond t$_{gel}$, the fractal dimension slowly increases up to $d_f \approx 2.3$, in full agreement with previous measurements on enzymatic milk gels under similar conditions~\citep{bauland2024}. Importantly, the change in abscissa of the break in slope at $q \simeq 3 \times 10^{-2}~\mathrm{nm^{-1}}$ is moderate throughout the entire gelation process, indicating that casein micelles remain well-defined entities of quasi-constant global size even in the gel state, in agreement with previous observations~\citep{li2018rheological}. In contrast, pronounced changes in the scattered intensity in regimes II and III point to hydrolysis-driven modifications of the internal structure of the casein micelles [Fig.~\ref{fig:gel}(c)--(d)].

\subsection{Progressive loss of casein micelle intermediate-q feature}
\label{sec:interm}

\subsubsection{Model-free analysis of the loss dynamics of intermediate-q feature}
\label{sec:modelfree}

\begin{figure}[t!]
    \includegraphics[scale=0.6, clip=true, trim=0mm 0mm 0mm 0mm]{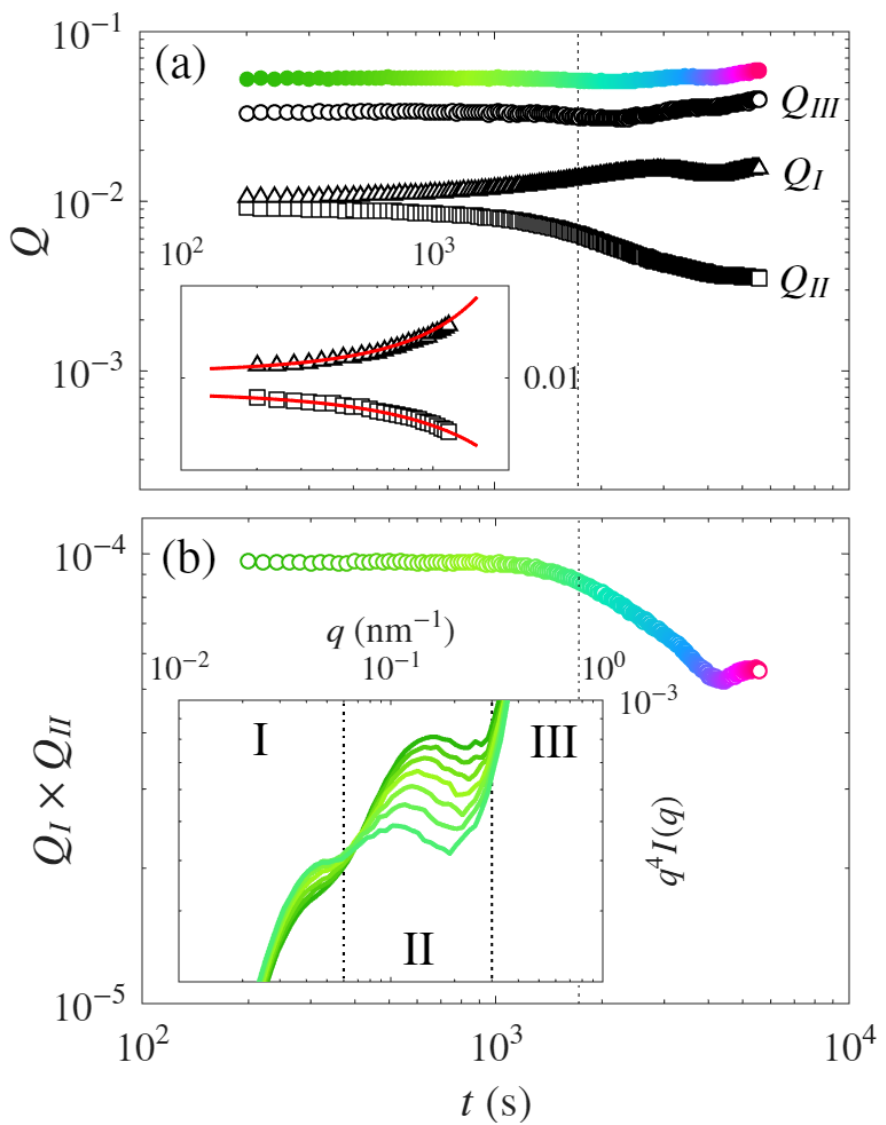}
    \centering
    \caption{(a) Temporal evolution of the scattering invariant $Q = \int_{0}^{\infty} I(q) q^2 \, dq$ (colored filled markers). Open markers correspond to windowed quantities termed ``partial integrals'' $Q_i$ (open markers), calculated over the low- (I), intermediate- (II), and high-q (III) regimes, as defined in Fig.~\ref{fig:gel}. Inset shows zoom on $Q_I$ and $Q_{II}$ vs. $t$. Red lines are the best exponential fits reading $Q_I = Q_{0} + A~\rm{exp}(t/\tau)$ and $Q_{II} = Q_{0} - A~\rm{exp}(t/\tau)$, respectively. (b) Temporal evolution of the product $Q_I \times Q_{II}$. Inset shows the time evolution of the SAXS spectra plotted as $q^4 I(q)$ vs. $q$, up to $t = 1000~\rm{s}$.
    }
    \label{fig:invariant}
\end{figure}

 The spectrum acquired shortly after enzyme addition [dark green curve in Fig.~\ref{fig:gel}(d)] exhibits a pronounced bump in the Kratky representation, centered at $q \simeq 0.15~\rm{nm}^{-1}$. Figure~\ref{fig:gel}(d) shows that this intermediary feature progressively vanishes following \emph{chymosin} addition and becomes almost undistinguishable at $t=t_{gel}$ (light blue curve indicated by an arrow). Accordingly, for $t$ > $t_{gel}$  (e.g., pink curve; $t = 5600~\rm{s}$) spectra display a power-law scaling $I(q) \propto q^{-4}$ in q-regime II, characteristic of a Porod regime associated with the form factor of highly polydisperse spheres with a smooth surface~\citep{li2016}. 

To quantify the kinetics associated with changes at different structural levels, we draw inspiration from the calculation of the scattering invariant $Q$ and compute ``partial integrals'' $Q_i = \int_{q_i}^{q_f} q^2 I(q)\, \mathrm{d}q$ for each previously defined q-regime. These quantities correspond to the area under the curve in the Kratky representation, allowing us to track intensity changes within each regime with higher precision as compared to tracking the temporal evolution of the intensity at a single $q$-value. Such a quantity has also been termed ``flocculation parameter'' when used to track the aggregation of nanoparticles~\citep{zhang2008}. Fig.~\ref{fig:invariant}(a) displays the temporal evolution of the partial integrals (empty symbols), together with the often called scattering invariant $Q$, corresponding to the total integral (filled circles). The latter, $Q = 2\pi^2 (\Delta\rho)^2 \phi (1-\phi)$~\citep{da2020,li2016}, depends only on the volume fraction of scatterers $\phi$ and the scattering contrast $\Delta\rho$, which are conserved during gelation, as reflected by the near-constant value of $Q$ over time. Similarly, the partial integral associated with the high-q regime, $Q_{III}$, remains approximately constant, indicating that the sum $Q_I$ + $Q_{II}$ is also approximately constant.

In contrast, the partial integrals $Q_I$ and $Q_{II}$ respectively increases and decreases from the earliest acquisition time following enzyme addition ($t = 200~\rm{s}$), as highlighted in the inset of Fig.~\ref{fig:invariant}(a). Notably, these changes occur well before the onset of casein micelle aggregation ($t \simeq 1000~\rm{s}$) as determined by DLS [Fig.~\ref{fig:gel}(b)], indicating that $\kappa$-casein hydrolysis alone modifies the scattering profile at both the overall casein micelle scale and the intermediate structural level. For better visualization of the scattering intensity change upon $\kappa$-casein hydrolysis and CM aggregation we present the product $Q_I \times Q_{II}$ in Fig.~\ref{fig:invariant}(b). Before significant aggregation, the product $Q_I \times Q_{II}$ remains constant up to $t \approx 1000~\rm{s}$. This behavior further evidences a correlation between the changes in scattering intensities occurring in regime I and regime II, up to the beginning of casein aggregation at $t = 1000~\rm{s}$. The concomitant increase and decrease of intensity occurring in regime I and II, respectively, is further highlighted in the inset of Fig.~\ref{fig:invariant}(b) which displays the evolution of $q^4 I(q)$ vs. $q$ over the timescale $t\leq 1000~\rm{s}$. Specifically, the intensity reduction of the intermediary bump in regime II occurs concomitantly with an increase in intensity of the colloid form factor contribution in regime I.

The anticorrelation between $Q_I$ and $Q_{II}$ is lost upon aggregation of casein micelles, as their product start decreasing for $t>1000~\rm{s}$. In the Porod regime, the scattered intensity per unit volume is dominated by surface scattering and reads $I(q) = 2\pi (\Delta\rho)^2 (S/V)\, q^{-4}$, where $S/V$ denotes the specific surface of the scatterers~\citep{li2016}. For $t > 1000~\rm{s}$, aggregation of destabilized casein micelles leads to a reduction of $S/V$, which further contributes to the decrease of scattering intensity in regime II. Accordingly, $Q_I \times Q_{II}$ decreases for $t > 1000~\rm{s}$, as the reduction of specific surface is not totally compensated by the formation of larger structures during aggregation (see Fig.~\ref{figsup:inva} in Appendix).

Beyond enzymatic hydrolysis, the amplitude of the intermediate-$q$ feature is known to depend more moderately on physicochemical conditions such as temperature and pH~\citep{pignon2004structure,sorensen2013characterisation,peyronel2020using,takagi2022temperature}. Two main interpretations have been proposed to account for this contribution: (i) scattering from a diffuse ``hairy'' layer of $\kappa$-casein surrounding an otherwise homogeneous casein matrix~\citep{shukla2009structure}, and (ii) the presence of an internal substructure with a characteristic size of $\approx 25~\rm{nm}$, composed of dense protein-rich regions, depicting casein micelles as porous objects~\citep{bouchoux2010squeeze,takagi2022temperature}. Such a heterogeneous internal structure was first inferred from osmotic compression experiments, which revealed the coexistence of dense, incompressible regions and softer, protein-depleted ones~\citep{bouchoux2010squeeze}.

The reduction of the intermediate feature following enzyme addition, depicted in Fig.~\ref{fig:gel}(d), initially appears consistent with the contribution of a $\kappa$-casein shell that would diminish as the enzymatic reaction proceeds. Within this framework, a direct relationship would be expected between the reduction of the intermediate feature and the rate of $\kappa$-casein hydrolysis at the colloid surface. However, closer examination reveals that at the onset of aggregation ($t \approx 1000~\rm{s}$), more than 85\% of $\kappa$-casein has been cleaved, while only about half of the scattering intensity associated with the intermediate bump has vanished [Fig.~\ref{fig:gel}(b) and Fig.~\ref{figsup:inva} in the Appendix]. Furthermore, the conservation of the scattering power, evidenced by the scattering invariant $Q$ remaining constant throughout enzymatic gelation, indicates that the hydrolyzed $\kappa$-casein fragment (N-terminal part) released into the whey does not contribute significantly to the SAXS signal. This is expected, as it is negatively charged and  highly hydrated under physiological conditions, providing limited scattering contrast. 

Accordingly, these results indicate that a simple loss of scattering from a $\kappa$-casein surface layer cannot fully account for the observed evolution of the SAXS profiles. The anticorrelated evolution of the scattering intensity in regimes I and II [Fig.~\ref{fig:invariant}(b)] instead suggests that $\kappa$-casein hydrolysis induces a redistribution of density gradients within the colloid. This points to a structural rearrangement from an initially heterogeneous substructure (the intermediate bump at $q \approx 0.15~\rm{nm}^{-1}$) toward a structure exhibiting reduced internal density fluctuations, consistent with a more homogeneous protein matrix characterized by a Porod regime $I(q) \propto q^{-4}$. If the intermediate-$q$ feature indeed reflects a porous architecture, the present results evidence that $\kappa$-casein plays a role in maintaining this internal structural organization which, to our knowledge, has not been previously reported. 

\subsubsection{Loss of the Intermediate-$q$ Feature Captured by Structural Models}

To challenge the microstructural scenario previously established, we quantitatively compare the enzyme-induced evolution of the spectra with the two structural models that have been proposed to account for the intermediate-$q$ feature: (i) a core-shell model describing a ``hairy'' layer of $\kappa$-casein surrounding a homogeneous casein matrix~\citep{shukla2009structure}, and (ii) a ``snowball'' model composed of two populations of spheres, where the smaller population describes dense, protein-rich regions, thereby depicting casein micelles as porous objects~\citep{sorensen2013characterisation,bouchoux2010squeeze}. The equations of both models are presented in detail in the~\ref{sec:sub:model}.

Because the objective of this section is to investigate the origin of the intermediate q-feature, the analysis is restricted to regimes I and II. The high-q contribution (regime III), generally attributed to the colloidal calcium phosphate nanoclusters, is not considered here and will be discussed separately in the following section. The purpose of these simulations is not to provide a unique solution for the kinetic pathway, but rather to assess whether physically plausible evolutions of each structural model can reproduce the experimentally observed redistribution of the scattering intensity. 

\begin{figure*}[t!]
    \centering
    \includegraphics[scale=0.4, clip=true, trim=0mm 0mm 0mm 0mm]{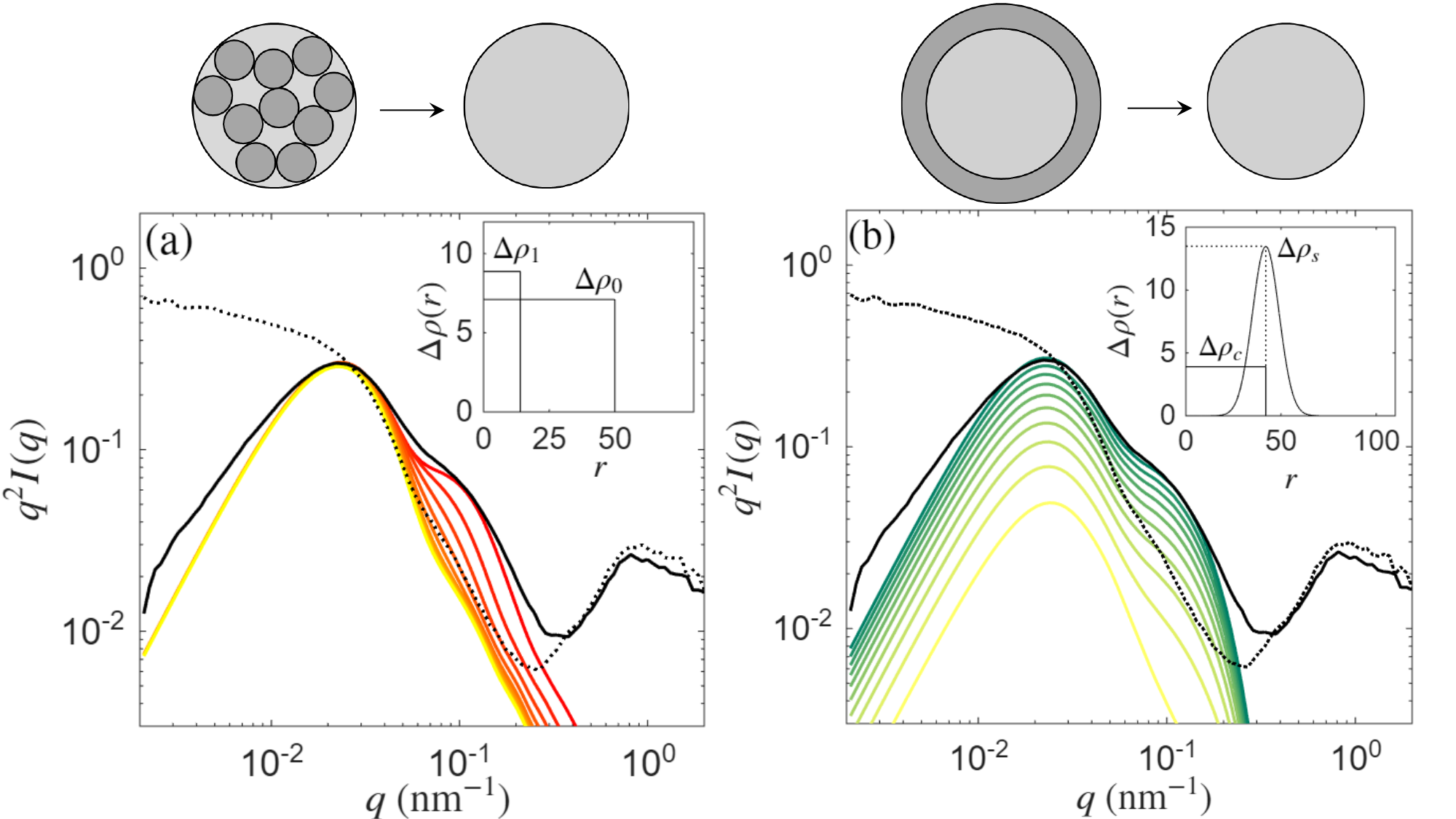}
    \caption{Evolution of the casein micelle scattering intensity upon enzyme addition captured by two structural models. (a) Time evolution (from red to yellow) captured by a hierarchical snowball model where the colloids ($r_0$) are composed of internal dense protein regions ($r_1$). (b) Time evolution (from green to yellow) captured by a diffuse core-shell model with a Gaussian density distribution for the shell ($\Delta \rho_s$). The plain and dotted black curves show the spectra of the liquid milk shortly after enzyme addition and of the aged gel, respectively. The insets of each panel display the scattering length densities ($\Delta \rho$) in $e/nm^3$ as a function of the particle radius ($r$). The schemes at the top of each panel depict the initial colloid structure evolving toward a homogeneous protein matrix.}
    \label{fig:model}
\end{figure*}

We first parametrize each model using the earliest spectrum of the kinetics [represented as a plain black line in Fig.~\ref{fig:model}(a)-(b)], corresponding to $t = 200~\rm{s}$ after enzyme addition. Both models initially capture the features of regimes I and II remarkably well. As shown in the inset of Fig.~\ref{fig:model}(a), the snowball model yields radii for the casein micelle and the dense protein regions of $r_0 = 50~\rm{nm}$ and $r_1 = 14~\rm{nm}$, respectively, with scattering length densities satisfying $\Delta \rho_1 > \Delta \rho_0$ in accordance with~\citep{bouchoux2010squeeze}. Similarly, for the core-shell model, the inset in Fig.~\ref{fig:model}(b) shows a casein micelle core radius of $r_c = 42~\rm{nm}$ and a shell thickness $t_s = 7~\rm{nm}$. As reported by~\citep{shukla2009structure}, the shell is assumed to be diffuse, following a Gaussian density profile, and denser than the core ($\Delta \rho_s > \Delta \rho_c$), a contrast difference required to reproduce the amplitude of the intermediate-$q$ feature observed in native casein micelles. Alternative low-contrast shells ($\Delta \rho_s < \Delta \rho_c$), explored in Fig.~\ref{figsup:model} of the Appendix, produce only a weak intermediate-$q$ contribution, although they may explain the slight increase in low-$q$ scattering intensity observed during hydrolysis.

From this initial point of the kinetics, we simulate the effect of enzymatic hydrolysis under a specific set of assumptions for each model. 
For the snowball model, we assume a relaxation of the porous structure into a homogeneous protein matrix by increasing the size of the dense regions until they reach the size of the overall casein micelle (i.e., $r_1 = r_0$). The number density of these dense protein regions is assumed to decrease with their growing size following a fractal scaling, such that the internal volume fraction of the dense regions is given by $\phi_{1_{\rm{int}}} = (r_0/r_1)^{d-3}$, where $d$ is the fractal dimension of the porous structure. Furthermore, the scattering length density of the dense regions decreases with their size according to $\Delta\rho_1 = \Delta\rho_0 (1 - \phi_{1_{\rm{int}}})/\phi_{1_{\rm{int}}}$, ensuring that $\Delta \rho_1 = \Delta \rho_0$ when $r_1 = r_0$.
For the core-shell model, we assume that the core-shell population is progressively converted into a pure core population—following $\phi_0 = \phi_{cs} + \phi_c$—meaning each colloid is treated either as an intact core-shell particle or as a naked core, since the shell is fully removed. 

The evolution of the simulated spectra for each model is compared to the final spectra of the kinetics, displayed as a dotted black line in Fig.~\ref{fig:model}(a)-(b). As shown in Fig.~\ref{fig:model}(a), the snowball model is fully consistent with the experimental dataset, capturing the progressive loss of the intermediate bump without any loss of scattering intensity at the casein micelle form factor. In contrast, for the core-shell model, the loss of the shell is concomitant with a decrease in intensity at the particle form factor. This reduction is attributed to the dominant contribution of the dense shell to the total scattering amplitude (see Fig.~\ref{figsup:model} in the Appendix for further details on the structural effect of the shell).

As suggested by the model-free analysis developed in Section~\ref{sec:modelfree}, the core-shell model is difficult to reconcile with the present dataset. Combined with the modeling framework presented here, we instead interpret the evolution of regime II as a relaxation of the internal porous structure of the colloid following chymosin addition. To our knowledge, these results provide the first evidence that $\kappa$-casein contributes not only to colloidal stabilization but also to the maintenance of the \emph{internal} structural organization of casein micelles. 

The presence of $\kappa$-casein within casein micelles had been suggested earlier based on proteolysis experiments with immobilized \emph{chymosin} (i.e. unable to penetrated inside the casein micelle), which showed a $13~\%$ reduction in $\kappa$-casein release compared to soluble \emph{chymosin}, consistent with reduced  $\kappa$-casein accessibility ~\citep{dalgleish1979proteolysis}. 
Because it carries a single phosphoserine residue, $\kappa$-casein has been regarded as a growth terminator during casein micelle assembly, owing to its weaker affinity for colloidal calcium phosphate, which accounts for its preferential localization at the micelle surface~\citep{horne1998}. In the recent quantitative multivalent binding model, however, $\kappa$-casein terminates the calcium–phosphate–mediated connectivity throughout the colloidal backbone, thereby generating an intermediate structural level and leading to its distribution across the entire micelle~\citep{holt2022quantitative,raynes2024structure}.

The increase in scattered intensity in regime~I at $q \simeq 3 \times 10^{-2}~\rm{nm^{-1}}$ shortly after enzyme addition [inset in Fig.~\ref{fig:invariant}(b)] could be explained by a slight swelling of the colloids resulting from the relaxation of casein micelle substructure. The scattered intensity associated with the casein micelle form factor reads $I(q) = N (\Delta \rho)^2 V^2 P(q)$, with $N$ the number density of scatterers and $P(q)$ the spherical form factor. Accordingly, $I(q) \propto V^2 \propto r^6$, implying that even small changes in micelle radius, hardly detectable directly in the spectra, can lead to significant variations in scattered intensity. An increase in scattered intensity could also results from the loss of a fluffy $\kappa$-casein shell from the surface of the casein micelle. While its contribution would be too weak to account for the intermediate-$q$ feature, $\kappa$-casein release from the colloid surface would locally increase the contrast. This is supported by a slight increase in intensity observed upon shell removal when the latter has a weak contrast as compare to the core, i.e., $\Delta \rho_s < \Delta \rho_c$ [see Fig.~\ref{figsup:model}(d) in the Appendix]. This pictures matches the general agreement that a fraction of $\kappa$-casein is located at the casein micelle surface, with its C-terminal part (caseinomacropeptide) being highly hydrated~\citep{holt1986electrophoretic,deKruif1996kappa,fox2008casein}. 

In summary, the progressive disappearance of the intermediate-$q$ feature reflects a relaxation of the casein micelle substructure and highlights the role of $\kappa$-casein located in the core of the casein micelle in maintaining its internal organization. Accordingly, the $\kappa$-casein location is not restricted to the colloid surface, but is present throughout the casein micelle. In the following section, we turn to the evolution of the scattering intensity in the high-$q$ regime (III).

\subsection{New peak in the high-q regime revealed in the gel state}
\label{sec:low}

\begin{figure*}[t!]
    \includegraphics[scale=0.35, clip=true, trim=0mm 0mm 0mm 0mm]{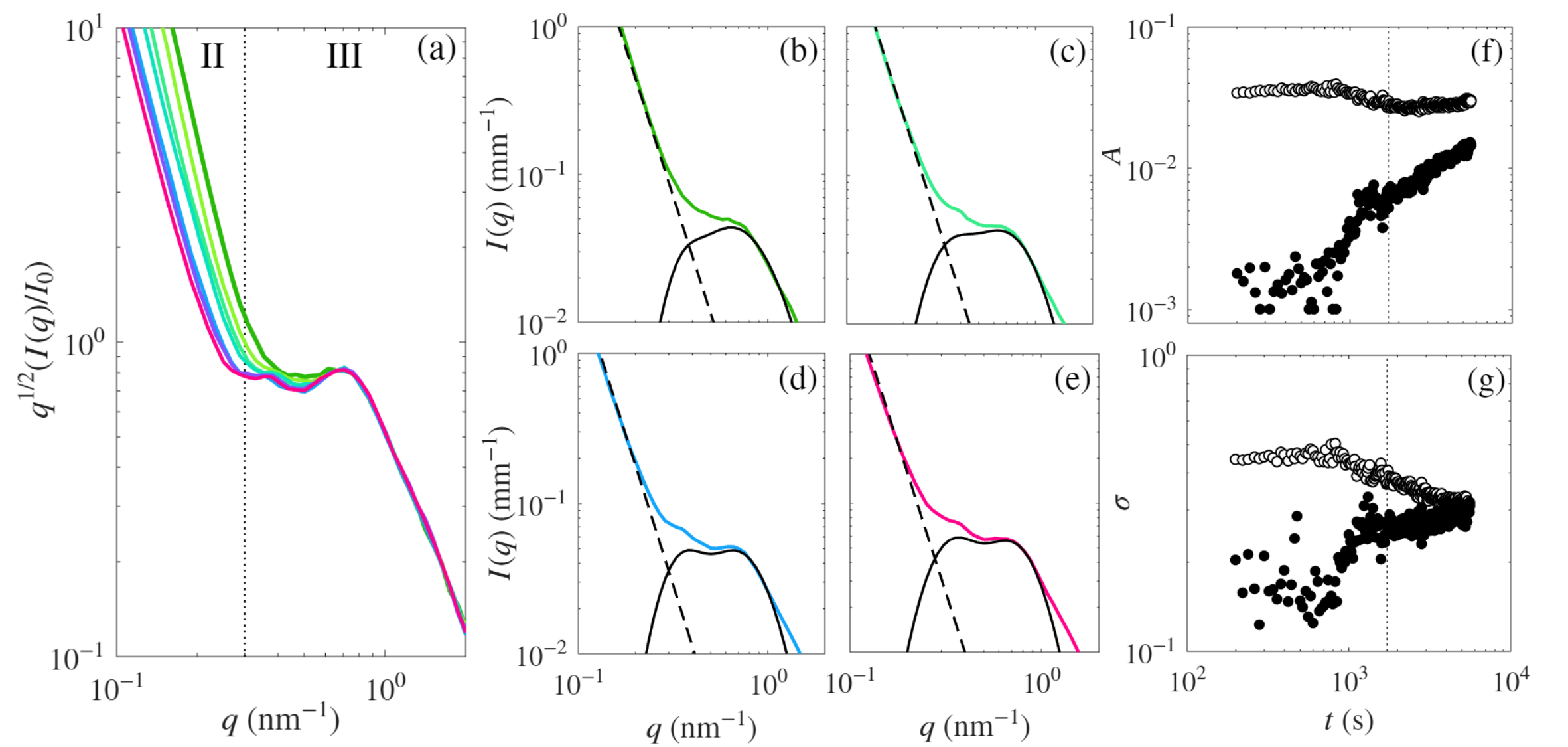}
    \centering
    \caption{(a) Temporal evolution (from green to pink) of the low-$q$ regime plotted as $q^{1/2} I(q)/I_0$ vs. $q$, where $I_0$ is the scattered intensity at $q = 0.7~\rm{nm}^{-1}$. (b)–(e) Best fits of the two peaks observed in the low-$q$ regime at $q = 0.4$ and $0.7~\rm{nm}^{-1}$ using two log-normal functions (solid black lines), after subtraction of the Porod contribution (dashed black lines). Panels (b), (c), (d), and (e) correspond to times $t = 200$, $1700$ ($t_{\mathrm{gel}}$), $3400$, and $5500~\rm{s}$, respectively. (f)–(g) Temporal evolution of the amplitude $A$ and width $\sigma$ of the log-normal functions. Filled and open symbols correspond to the peaks at $q = 0.4$ and $0.7~\rm{nm}^{-1}$, respectively.
    }
    \label{fig:highq}
\end{figure*}

Fig.~\ref{fig:highq}(a) shows the temporal evolution of the scattered intensity in the high-q regime, normalized by the intensity at $q = 0.7~\rm{nm}^{-1}$ and multiplied by $q^{1/2}$ to enhance visualization [see Fig.~\ref{fig:highq}(a) and Fig.~\ref{figsup:highq} in Appendix for unnormalized data]. 

Focusing on the initial spectrum in Fig.~\ref{fig:highq}(a) (dark green curve; $t = 200~\rm{s}$), the high-q regime exhibits a peak centered at $q = 0.7~\rm{nm}^{-1}$,  attributed to small-scale density fluctuation of casein chains \citep{ingham2016revisiting} or calcium phosphate nanoclusters \citep{bouchoux2010squeeze, takagi2022temperature} or both \citep{de2012casein, pedersen2022model}. Note that a similar peak, centered at $q \approx 0.67~\rm{nm}^{-1}$, was also observed for microgels of milk serum proteins, i.e., $\alpha$-lactalbumin and $\beta$-lactoglobulin, where salt precipitates are absent~\citep{schmitt2010}. For $q > 0.7~\rm{nm}^{-1}$, the intensity decays as $I(q) \propto q^{-2.2}$, consistent with a coil-like conformation of casein chains \citep{marchin2007effects, pedersen2022model}. While these features remain essentially unchanged during enzymatic gelation, the progressive reduction of scattered intensity in the intermediary regime leads to the emergence of a second peak centered at $q = 0.4~\rm{nm}^{-1}$ in agreement with \citep{li2018rheological} [see, for instance, the pink curve in Fig.~\ref{fig:highq}(a)]. This peak has previously been observed in small-angle neutron scattering experiments performed at the protein contrast-matching point, and was consequently attributed to \rev{an interparticle} structure factor\rev{, brought about through the organization of density fluctuation around the} colloidal calcium phosphate nanoclusters \citep{stothart1982small, holt2003substructure}. It was also observed by SAXS in enzymatic milk gels prepared in D\textsubscript{2}O~\citep{li2018rheological} or after increasing the volume fraction of the casein micelles such as in dried milk powders \citep{mata2011structure} and at the surface of an ultrafiltration membrane \citep{pignon2004structure}. 

To quantify the temporal evolution of these two peak features, we fit them using lognormal functions of the form
$G_i(q) = \frac{A_i}{\sqrt{2\pi}\sigma_i q}\exp\left[-\frac{\ln^2(q/q_i)}{2\sigma_i^2}\right]$ where $q_i$ denotes the peak position, $\sigma_i$ its width, and $A_i$ its amplitude. Here, $q_i$ is fixed at $0.4$ and $0.7~\rm{nm}^{-1}$. Fig.~\ref{fig:highq}(b)--(e) show the fits performed after subtraction of a power-law contribution $I(q) = I_0 q^{-4}$ corresponding to the Porod regime of the casein micelle form factor (see section~\ref{sec:sub:highq} in Appendix for details).

Fig.~\ref{fig:highq}(f)--(g) display the temporal evolution of the amplitudes $A$ and widths $\sigma$ of the two peaks. Focusing first on the peak at $q = 0.7~\rm{nm}^{-1}$ (empty symbols), its amplitude slightly decreases around $t \approx 1000~\rm{s}$. \rev{A slight decrease in scattering intensity is observed over a broader $q$ range around and after gelation [Fig.~\ref{fig:invariant}(a)]. Radiation damage can induce aggregation and low-$q$ artifacts in protein SAXS
measurements~\citep{bouchoux2010squeeze,nogueira2020multiscale}. Although the short and intermittent exposures used here (100~ms every 20~s) were chosen to reduce cumulative irradiation, a minor beam-induced contribution cannot be excluded at late times, when gelation restricts diffusive renewal of the illuminated volume. Nevertheless, regime III remains essentially constant during the pre-gelation stage, and the dominant temporal changes are strongly $q$-dependent and consistent with the structural evolution associated with enzymatic gelation. Possible radiation damage is therefore unlikely to affect the principal conclusions.} The onset of aggregation at $t \approx 1000~\rm{s}$ also coincides with an increase in the amplitude of the peak at $q = 0.4~\rm{nm}^{-1}$ (filled symbols), as the Porod contribution in regime II shifts toward lower $q$ values. For $t > 1000~\rm{s}$, the further increase of the $0.4~\rm{nm}^{-1}$ peak amplitude primarily reflects the decrease of Porod scattering intensity in the intermediary regime (see Fig.~\ref{figsup:inva} in Appendix). Indeed, structural aging in enzymatic milk gels has been shown to proceed via contact aging between particles~\citep{Mellema2002,bauland2024}, likely involving partial fusion of casein micelles and a further reduction of the specific surface. While the final spectra corresponding to aged gels are nearly superimposed at $q = 0.4~\rm{nm}^{-1}$ [Fig.~\ref{fig:highq}(a)], the simplified fitting procedure used here remains sensitive to subtle changes in the Porod regime at intermediate $q$, which artificially produces a slight increase in the fitted peak amplitude at late times.

In summary, we show that the reduction of the specific surface area of casein micelles during aggregation reveals a peak centered at $q = 0.4~\text{nm}^{-1}$. This feature, traditionally observed using small-angle neutron scattering at the protein contrast matching point, is attributed to the structure factor of calcium-phosphate nanoclusters. In agreement with earlier results obtained in D~\textsubscript{2}0, we show that in the gel state, this feature is easily detectable via X-ray scattering even at low volume fractions and without any solvent modification. Importantly, the emergence of the peak at  $q = 0.4~\text{nm}^{-1}$ does not necessarily imply the formation of a new structural motif during gelation, but rather the progressive unmasking of a pre-existing correlation feature as the Porod contribution associated with the casein micelle interface decreases. The approximately constant peak position throughout gelation further supports its assignment to an intrinsic structural length scale, consistent with previous SANS studies attributing this feature to correlations between colloidal calcium phosphate nanoclusters. These results demonstrate that a structural feature classically accessed by contrast-matching SANS can become directly accessible by SAXS when the Prorod contribution is sufficiently reduced. The significantly faster acquisition times offered by X-ray scattering open opportunities for the straightforward characterization of the internal structure of casein micelles under varying physicochemical conditions, provided that the convolution with Porod scattering is properly accounted for.

\section{Conclusion} 

We have shown that the enzymatic gelation of milk provides a unique experimental window into the hierarchical organization of casein micelles. By combining time-resolved SAXS measurements with complementary analyses of $\kappa$-casein hydrolysis, particle aggregation, and gel formation, we established temporal correlations between enzymatic cleavage and structural rearrangements occurring at multiple length scales during the enzymatic sol-gel transition of milk. Based on structural models, our results demonstrate that beyond its well-established role as a steric stabilizer and growth terminator, $\kappa$-casein is fundamental in maintaining the internal structural integrity of the colloid, as evidenced by the progressive relaxation of the intermediate-$q$ level following enzymatic cleavage. Furthermore, we highlight that the reduction of the specific surface area between the colloids and their surrounding solvent in the gel state unmasks the structural signature of calcium-phosphate nanoclusters, offering a straightforward route for their characterization using X-ray scattering. Finally, this work underscores the advantage of combining small-angle scattering experiments with complementary techniques, such as dynamic light scattering and rheometry, as it provides the essential temporal references required to interpret complex scattering patterns and their dynamics in evolving soft-matter systems.

\section*{Conflicts of interest}
There are no conflicts to declare.

\section*{Data availability} 
The data that support the findings of this study are available from the corresponding author upon reasonable request.

\section*{Acknowledgments}

The authors acknowledge the European Synchrotron Radiation Facility (ESRF) for financial support and for the provision of synchrotron radiation facilities under proposal number SC-5658 at beamline ID02. The authors are grateful to Theyencheri Narayanan for technical support during the synchrotron experiments and usefull discussions. Erik Juste and Fanny Guyomarc'h are also acknowledged for experimental assistance at ESRF, as well as Cecilia Candido Rodrigues de Souza and Marina Cyrillo Gusella for experimental assistance during enzymatic hydrolysis, dynamic light scattering, and rheology analysis. We also acknowledge the research group CNRS-INRAE ``GDR SLAMM'' for helpful discussions on scattering data interpretation.

\clearpage 

\appendix
\setcounter{figure}{0}
\setcounter{table}{0}
\setcounter{equation}{0}
\label{sec:app}

\section{Preparation and Choice of Background}
\label{sec:sub:bck}

The background for SAXS measurements was the serum phase of the milk after gelation. To recover the serum, the enzymatic milk gel was cut into small pieces to induce syneresis. The resulting curd/whey mixture was centrifuged at 14,000~g for 20~min at 20$^{\circ}$C, and the supernatant was filtered through a 0.1~$\mu$m filter to recover the filtrate (the serum phase). For comparison, a serum phase was also produced from native milk (without enzyme addition) by ultracentrifugation at 100,000~g for 1~h, followed by identical filtration.

\begin{figure}
    \includegraphics[scale=0.55, clip=true, trim=0mm 0mm 0mm 0mm]{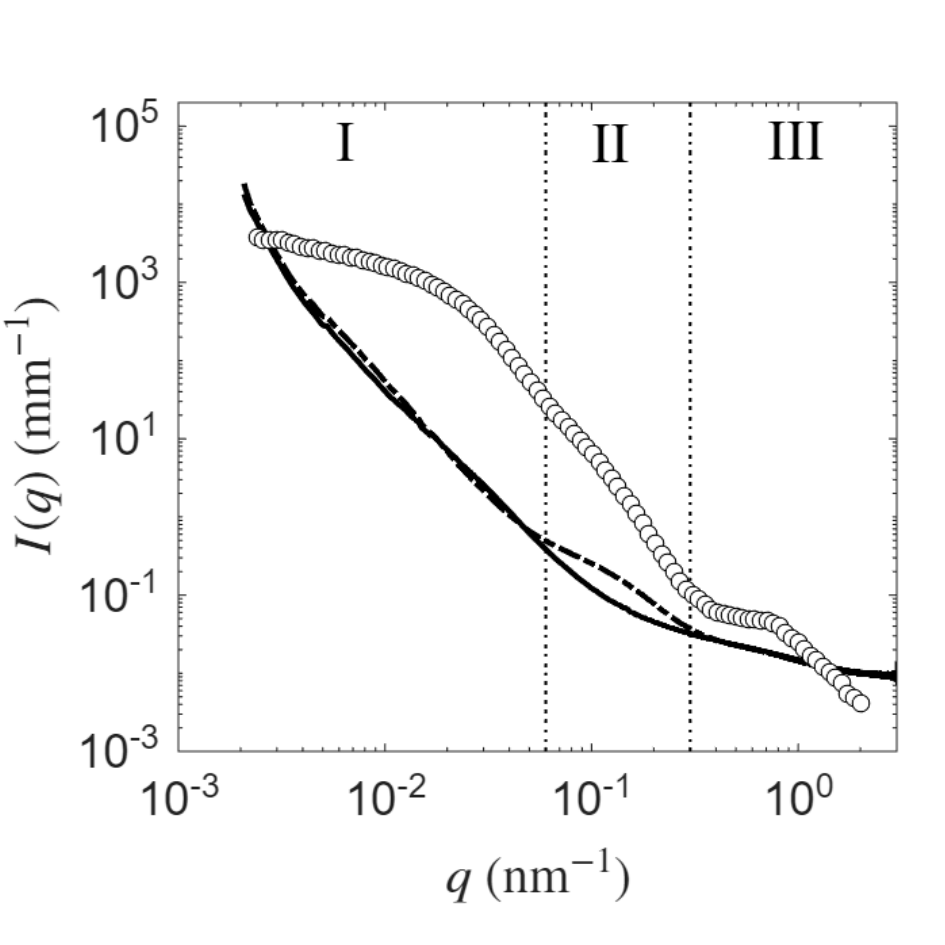}
    \centering
    \caption{Scattered intensity $I(q)$ of the supernatant obtained by centrifugation before (dotted line) and after (solid line) enzymatic gelation. The empty markers represent the scattered intensity of the milk immediately after enzyme addition ($t = 200~\text{s}$), using the gel supernatant for background subtraction.
    }
    \label{figsup:bck}
\end{figure}

Fig.~\ref{figsup:bck} compares the scattering of the native milk supernatant (dotted black line) and the gel supernatant (solid black line). While the scattered intensity of these backgrounds is two orders of magnitude lower than that of the native milk (empty markers), the milk supernatant exhibits higher scattering in the intermediate-$q$ range compared to the gel supernatant. Such a contribution could be interpreted as very small casein micelles or protein assemblies remaining in the solvent. However, based on neutron scattering experiments at the protein matching point, previous reports have attributed this intermediate feature to residual phospholipids in the milk~\citep{bouchoux2015}. In the gel state, these lipids would be trapped within the protein network, explaining their absence in the gel supernatant. Consequently, we selected the gel supernatant as the reference background. Given the high scattering intensity of the milk samples in this range, the structural results discussed in the manuscript remain unaffected by the choice of background.

\section{Modeling of Low and Intermediate-$q$ Features During Enzymatic Hydrolysis}
\label{sec:sub:model}

\begin{figure}[t!]
    \centering
    \includegraphics[scale=0.55, clip=true, trim=0mm 0mm 0mm 0mm]{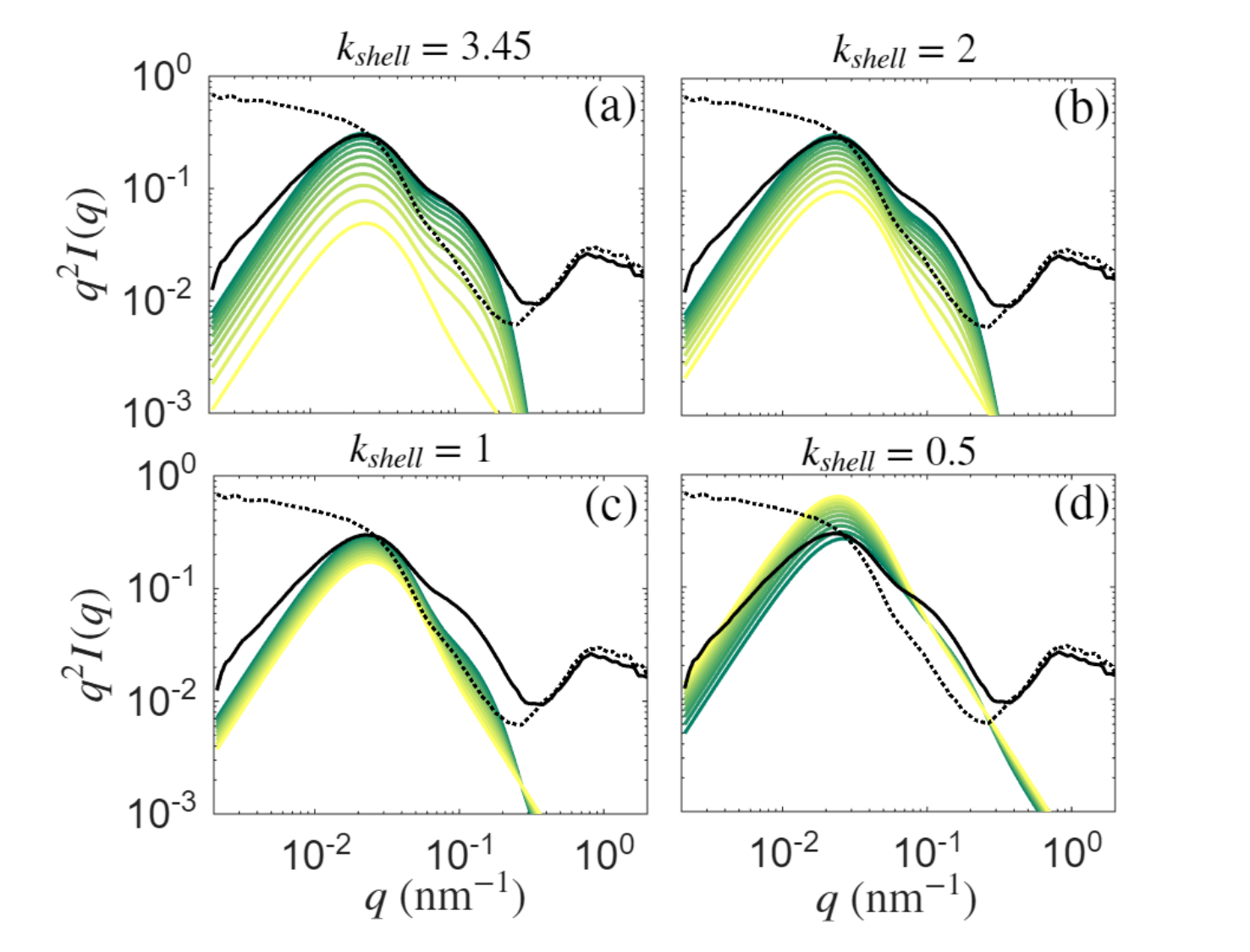}
    \caption{Effect of the relative contrast of the shell $k_{shell} = \Delta \rho_s / \Delta \rho_c$ for the diffuse core-shell model. From green to yellow: transition from a population of core-shell to pure core particles.}
    \label{figsup:model}
\end{figure}

We aim to confront our experimental dataset with the two structural models reported to account for the intermediate-$q$ feature in casein micelle scattering spectra. As presented in the main text, these two models consist of (i) a \emph{core-shell model} describing a ``hairy'' layer of $\kappa$-casein surrounding a homogeneous casein matrix~\citep{shukla2009structure}, and (ii) a \emph{hierarchical snowball model} composed of two populations of spheres, where the smaller population describes dense, protein-rich regions~\citep{bouchoux2010squeeze}. 

We begin by detailing both models following the frameworks established by Shukla et al.~\citep{shukla2009structure} and Bouchoux et al.~\citep{bouchoux2010squeeze}.

\subsection{Hierarchical Snowball Model}

The model proposed by Bouchoux et al.~\citep{bouchoux2010squeeze} consists of an assembly of small spheres nested within a larger sphere, assuming no correlation terms between the distinct structural levels. Consequently, the total scattered intensity is expressed as the sum of the contributions from the overall casein micelles (level 0) and the internal dense regions (level 1):

\begin{align}
    I(q) & = I_0(q) + I_1(q) \\
    I(q) & = \phi_0 \Delta\rho_0^2 V_0 G(q, r_0)^2 + \phi_1 \Delta\rho_1^2 V_1 G(q, r_1)^2
\end{align}

\noindent where $r_0$ and $r_1$ are the radii of the large and small spheres, with corresponding volumes $V_0$ and $V_1$, and $\Delta\rho_0$ and $\Delta\rho_1$ represent their respective scattering length density contrasts. $G(q, r)$ is the normalized spherical form factor amplitude, defined as:

\begin{equation}
    G(q, r) = \frac{3[\sin(qr) - qr\cos(qr)]}{(qr)^3}
\end{equation}

Unlike the original work by Bouchoux et al.~\citep{bouchoux2010squeeze} which implemented a Schulz distribution, polydispersity is accounted for here via numerical integration over a log-normal size distribution function:

\begin{equation}
    N(r) = \frac{1}{r\sigma \sqrt{2\pi}}\exp\left[-\frac{1}{2}\left(\frac{\ln(r/r_i)}{\sigma}\right)^2 \right]
\end{equation}

\noindent where $r_i$ is the median radius associated with level $i$ and $\sigma$ is the mid-width parameter of the distribution. We set $\sigma = 0.3$ for both levels.

\subsection{Diffuse Core-Shell Model}

To account for casein micelle polydispersity alongside a Gaussian density profile of the $\kappa$-casein layer, Shukla et al.~\citep{shukla2009structure} formulated the diffuse core-shell model as a sum of the shell ($A_s$) and core ($A_c$) contributions to the total scattering amplitude:

\begin{align}
    I_{\rm{cs}}(q) &= \frac{\phi_0}{V(r_c)}(A_c(q) + A_s(q))^2 \\
    A_c(q) &= \Delta \rho_c \frac{4 \pi}{q^3} \left[\sin(qr_c)-qr_c\cos(qr_c)\right] \\
    A_s(q) &= \Delta \rho_s \frac{4 \pi\sqrt{2\pi}t_s}{q} \exp\left(-\frac{q^2t_s^2}{2}\right) \left[qt_s^2 \cos(qr_c) + r_c\sin(qr_c)\right]
\end{align}

\noindent where $r_c$ is the core radius, $t_s$ is the shell thickness, and $\Delta \rho_c$ and $\Delta \rho_s$ are the respective scattering length density differences relative to the solvent. 

To account for structural polydispersity, the total scattering intensity is numerically integrated over a log-normal distribution function assigned to the core radius $r_c$, while maintaining a constant shell thickness $t_s$.

\subsection{Modeling Approach and Kinematics Simulation}

To assess the effect of hydrolysis within these two structural models, we first parametrize each model using the initial scattering spectra of the kinetics, corresponding to native casein micelles. The total volume fraction of the casein micelles is fixed at $\phi_0 = 0.1$, which matches values expected from casein concentrations close to physiological milk conditions~\citep{bauland2024}. 

As summarized in Table~\ref{tab:model}, the initial parametrization yields values in good agreement with literature data. Specifically, for the snowball model, the size of the dense protein regions ($r_1 = 14~\rm{nm}$ compared to the reported $10~\rm{nm}$) and their relative volume fraction closely match the values obtained by Bouchoux et al.~\citep{bouchoux2010squeeze}. Similarly, for the core-shell model, the calculated shell thickness ($t_s = 7~\rm{nm}$ compared to the reported $12~\rm{nm}$) is consistent with the value given by Shukla et al.~\citep{shukla2009structure}.

\begin{table}[h!]
\centering
\caption{Comparison of the models parametrization with the literature}
\label{tab:model}
\begin{tabular}{lllllll}
\toprule
\textbf{Snowball} & $\phi_0$ & $\Delta\rho_0$ ($\text{e/nm}^3$) & $r_0$ (nm) & $\phi_1$ & $\Delta\rho_1$ & $r_1$ (nm) \\
\midrule
Bouchoux et al.~\citep{bouchoux2010squeeze} & 0.11 & 15 & 46 & 0.055 & 16 & 10 \\
Present work    & 0.1  & 7  & 50 & 0.053 & 9  & 14 \\
\midrule
\textbf{Core-shell} & $\phi_0$ & $\Delta\rho_c$ ($\text{e/nm}^3$) & $r_c$ (nm) & $\Delta\rho_s$ ($\text{e/nm}^3$) & $t_s$ (nm) & \\
\midrule
Shukla et al.~\citep{shukla2009structure}   & 0.1  & 14 & 55 & 25* & 12 & \\
Present work    & 0.1  & 4  & 42 & 13  & 7  & \\
\bottomrule
\end{tabular}
\end{table}

Following this initialization step, we keep most of the parameters constant and simulate the effect of enzymatic hydrolysis under a specific set of assumptions for each model.

For the snowball model, we assume a relaxation of the porous structure into a homogeneous protein matrix by increasing the size of the dense regions until they reach the size of the overall casein micelle (i.e., $r_1 = r_0$). The number density of these dense protein regions is assumed to decrease with their growing size following a fractal scaling, such that the internal volume fraction of the dense regions is given by:

\begin{align}
    \phi_{1_{\rm{int}}} & = (r_0/r_1)^{d-3} \\
    \phi_1 & = \phi_0 \times \phi_{1_{int}}
\end{align}

\hspace{1pt} where $d$ is the fractal dimension of the porous structure. Here, we find $d=2.5$, yielding $\phi_0/\phi_1 \approx 0.5$. Furthermore, as proposed by Bouchoux et al. in the context of osmotic compression~\citep{bouchoux2010squeeze}, the scattering length density of the dense regions decreases with their size according to:

\begin{equation}
    \Delta\rho_1 = \Delta\rho_0 \frac{1 - \phi_{1_{\rm{int}}}}{\phi_{1_{\rm{int}}}}
\end{equation}

This boundary condition naturally ensures that the contrasts equalize ($\Delta \rho_1 = \Delta \rho_0$) when full homogenization is reached ($r_1 = r_0$).

For the core-shell model, the enzymatic reaction is assumed to progressively convert the core-shell population into a pure core population. The total volume fraction is preserved via $\phi_0 = \phi_{\rm{cs}} + \phi_c$, meaning each colloid is treated either as an intact core-shell particle or as a bare core with its hairy layer fully removed. The resulting scattered intensity is modeled as the linear combination of both populations:

\begin{equation}
    I(q) = \frac{\phi_{\rm{cs}}}{V(r_{\rm{cs}})}\langle(A_c(q) + A_s(q))^2\rangle + \frac{\phi_c}{V(r_c)}\langle A_c(q)^2\rangle
\end{equation}

\noindent where at $t=0$, $\phi_c = 0$, and as the reaction approaches completion ($t \to \infty$), $\phi_{\rm{cs}}(t) \to 0$.

In the main text, we show that the progressive loss of the shell is associated with a decrease of intensity at the level of the casein micelle form factor, which is not seen in the data. This is attributed to the strong scattering of the dense shell. Defining $k_{\rm{shell}} = \Delta \rho_s / \Delta \rho_c$ as the ratio of the scattering length density differences of the shell and core, we show in Fig.~\ref{figsup:model} the effect of shell removal on the scattering spectra for different $k_{\rm{shell}}$ values. Setting $k_{\rm{shell}} = 3.45$ as done in Fig.~\ref{figsup:model} and Fig.~\ref{fig:model}(b) in the main text is necessary to properly capture the intermediate-$q$ feature. Decreasing $k_{\rm{shell}}$ reduces the loss of scattering intensity at low-$q$ during the shell removal, but is unable to properly capture the intermediate-$q$ feature of the native casein micelles.

\section{Fit of the high-q regime with log-normal functions}
\label{sec:sub:highq}
\begin{figure}
    \includegraphics[scale=0.55, clip=true, trim=0mm 0mm 0mm 0mm]{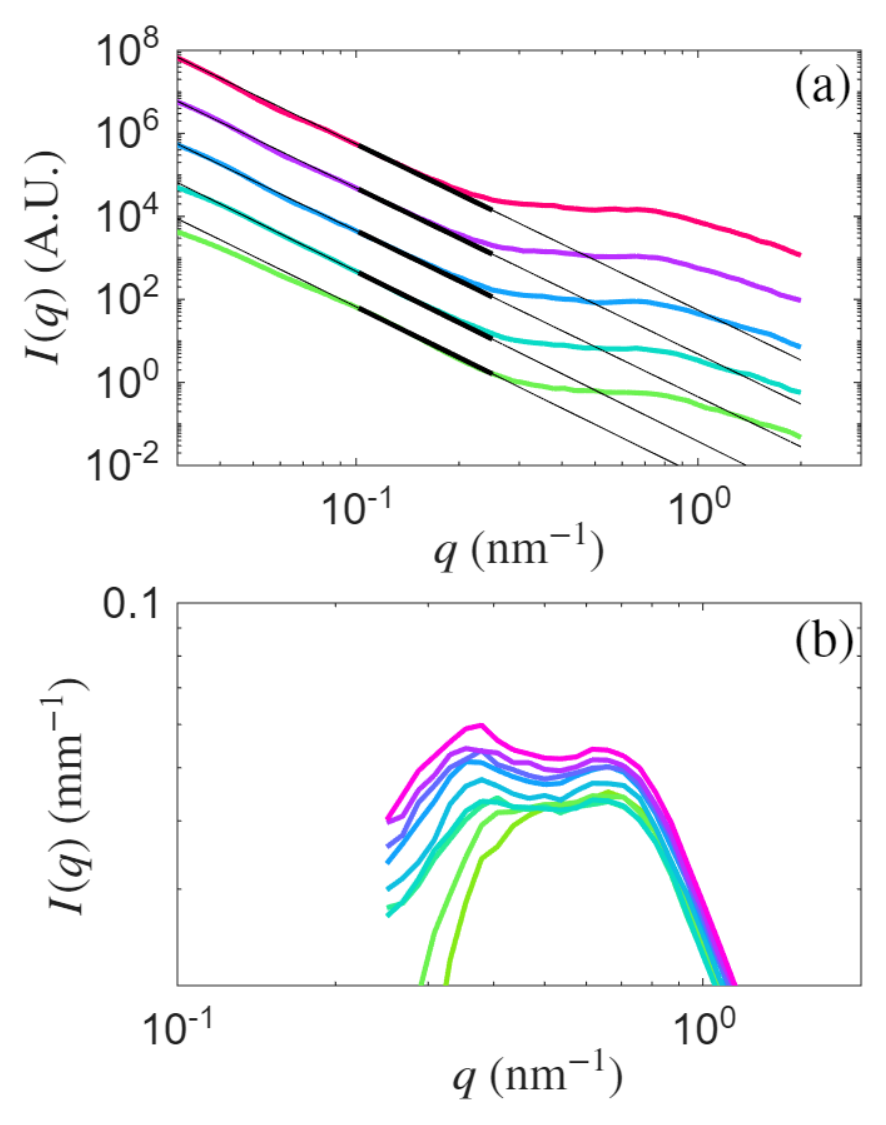}
    \centering
    \caption{(a) Fit of the SAXS data with a Porod model $I(q) = I_0q^{-4}$. Thick and thin black lines are the actual fit, performed over a fixed interval $0.105 < q < 0.125~\rm{nm}^{-1}$, and its extrapolation over the full q-range, respectively. (b) Corresponding spectra after subtraction of the Porod regime described in (a). 
    }
    \label{figsup:fit}
\end{figure}

As described in the main text, the two peaks visible in regime III are fit using log-normal functions after the subtraction of a power-law contribution, $I(q) = I_0 \times q^{-4}$, corresponding to the Porod regime of the casein micelle form factor. For each spectrum, the parameter $I_0$ is determined by fitting the data in the range $0.105 < q < 0.125~\rm{nm}^{-1}$ prior to subtraction [Fig.~\ref{figsup:fit}(a)]. Fig.~\ref{figsup:fit}(b) displays the resulting data after the power-law background has been removed.

\section{Additional representation}
\begin{figure}
    \includegraphics[scale=0.55, clip=true, trim=0mm 0mm 0mm 0mm]{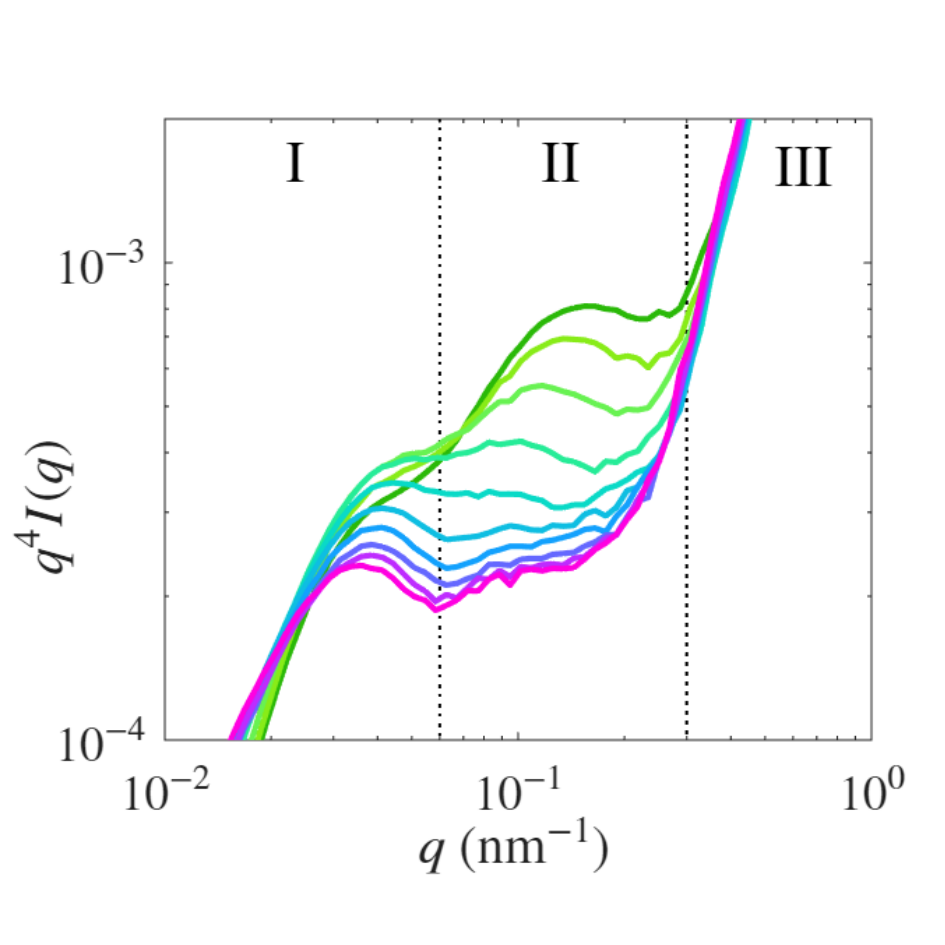}
    \centering
    \caption{Focus on q-regime I and II defined in the main text. Colors code for the time evolution of the SAXS spectra, plotted as $q^4 I(q)$ vs. $q$. From top to bottom, $t =200$, 740, 1280, 1820, 2360, 2900, 3440, 3980 and $4520~\rm{s}$.
    }
    \label{figsup:inva}
\end{figure}

\begin{figure}
    \includegraphics[scale=0.55, clip=true, trim=0mm 0mm 0mm 0mm]{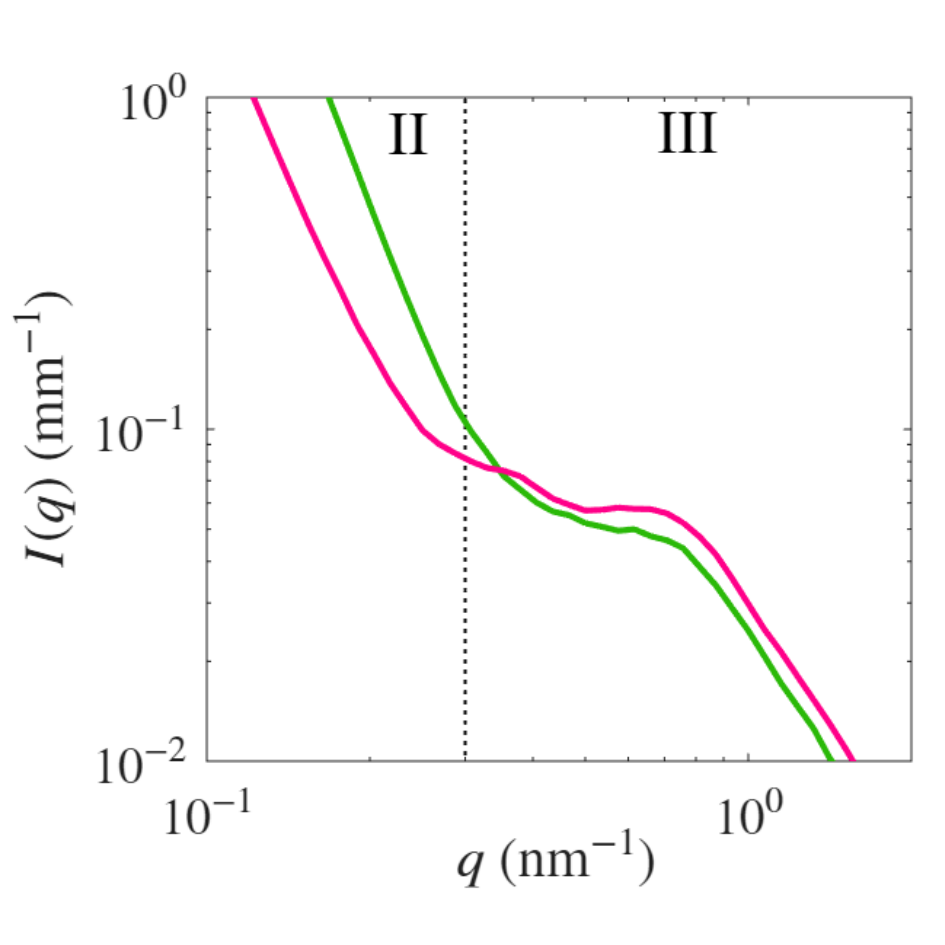}
    \centering
    \caption{Focus on q-regime III defined in the main text. Green and pink curves correspond to the scattered intensity $I(q)$ vs $q$ measured at $t=200$ and $t=5600~\rm{s}$.
    }
    \label{figsup:highq}
\end{figure}

Fig.~\ref{figsup:inva} provides an additional visualization of the SAXS data, $q^4I(q)$ vs $q$, as referred to in the main text. Fig.~\ref{figsup:highq} provides a detailed view of the high-$q$ region for the unrescaled data, complementing the rescaled representation shown in Fig.~\ref{fig:highq}(a) of the main text.

\clearpage


\bibliographystyle{elsarticle-num}
\bibliography{library}

@article{deKruif1996kappa,
  title={$\kappa$-casein as a polyelectrolyte brush on the surface of casein micelles},
  author={De Kruif, CG and Zhulina, Ekatherina B},
  journal={Colloids and Surfaces A: Physicochemical and Engineering Aspects},
  volume={117},
  number={1-2},
  pages={151--159},
  year={1996},
  publisher={Elsevier}
}

@article{holt1986electrophoretic,
  title={Electrophoretic and hydrodynamic properties of bovine casein micelles interpreted in terms of particles with an outer hairy layer},
  author={Holt, Carl and Dalgleish, Douglas G},
  journal={Journal of Colloid and Interface Science},
  volume={114},
  number={2},
  pages={513--524},
  year={1986},
  publisher={Elsevier}
}

@article{moller2025structural,
  title={Structural changes of calcium-depleted casein micelle suspensions studied by SAXS},
  author={M{\o}ller, Thea Lykkegaard and Raak, Norbert and Pedersen, Jan Skov and Corredig, Milena},
  journal={Colloids and Surfaces A: Physicochemical and Engineering Aspects},
  volume={711},
  pages={136274},
  year={2025},
  publisher={Elsevier}
}

@article{zhang2008,
  title={Gold nanoparticles decorated with oligo (ethylene glycol) thiols: kinetics of colloid aggregation driven by depletion forces},
  author={Zhang, Fajun and Dre{\ss}en, Donald G. and {\v{S}}koda, Maximilian W. A. and Jacobs, Robert M. J. and Zorn, Stefan and Martin, Richard A. and Martin, Christopher M. and Clark, Graham F. and Schreiber, Frank},
  journal={European Biophysics Journal},
  volume={37},
  number={5},
  pages={551--561},
  year={2008},
  publisher={Springer}
}

@article{Mellema2002,
author = {Mellema, M. and Walstra, P. and {Van Opheusden}, J. H.J. and {Van Vliet}, T.},
doi = {10.1016/S0001-8686(01)00089-6},
issn = {00018686},
journal = {Adv. Colloid Interface Sci.},
month = {apr},
number = {1},
pages = {25--50},
pmid = {12061711},
publisher = {Elsevier},
title = {{Effects of structural rearrangements on the rheology of rennet-induced casein particle gels}},
volume = {98},
year = {2002}
}

@article{Begam2021,
title = {Kinetics of Network Formation and Heterogeneous Dynamics of an Egg White Gel Revealed by Coherent X-Ray Scattering},
  author = {Begam, Nafisa and Ragulskaya, Anastasia and Girelli, Anita and Rahmann, Hendrik and Chandran, Sivasurender and Westermeier, Fabian and Reiser, Mario and Sprung, Michael and Zhang, Fajun and Gutt, Christian and Schreiber, Frank},
  doi = {10.1103/PhysRevLett.126.098001},
  journal = {Phys. Rev. Lett.},
  number = {9},
  volume = {126},
  pages = {098001},
  year = {2021},
  publisher = {American Physical Society}
}

@article{zaccone2010,
  title={Shear-induced reaction-limited aggregation kinetics of Brownian particles at arbitrary concentrations},
  author={Zaccone, Alessio and Gentili, Daniele and Wu, Hua and Morbidelli, Massimo},
  journal={The Journal of chemical physics},
  volume={132},
  number={13},
  year={2010},
  publisher={AIP Publishing}
}

@article{li2016,
  title={Small angle X-ray scattering for nanoparticle research},
  author={Li, Tao and Senesi, Andrew J and Lee, Byeongdu},
  journal={Chemical reviews},
  volume={116},
  number={18},
  pages={11128--11180},
  year={2016},
  publisher={ACS Publications}
}

@article{schmitt2010,
author = {Schmitt, Christophe and Moitzi, Christian and Bovay, Claudine and Rouvet, Martine and Bovetto, Lionel and Donato, Laurence and Leser, Martin E and Schurtenberger, Peter and Stradner, Anna},
doi = {10.1039/c0sm00220h},
issn = {1744683X},
journal = {Soft Matter},
number = {19},
pages = {4876--4884},
title = {{Internal structure and colloidal behaviour of covalent whey protein microgels obtained by heat treatment}},
volume = {6},
year = {2010}
}

@article{bouchoux2015,
  title={Structural heterogeneity of milk casein micelles: a SANS contrast variation study},
  author={Bouchoux, Antoine and Ventureira, Jorge and G{\'e}san-Guiziou, Genevi{\`e}ve and Garnier-Lambrouin, Fabienne and Qu, Peng and Pasquier, Coralie and P{\'e}zennec, St{\'e}phane and Schweins, Ralf and Cabane, Bernard},
  journal={Soft Matter},
  volume={11},
  number={2},
  pages={389--399},
  year={2015},
  publisher={Royal Society of Chemistry}
}

@article{horne1998,
  title={Casein interactions: casting light on the black boxes, the structure in dairy products},
  author={Horne, David S},
  journal={International Dairy Journal},
  volume={8},
  number={3},
  pages={171--177},
  year={1998},
  publisher={Elsevier}
}

@article{da2020,
  title={Interplay between glass formation and liquid--liquid phase separation revealed by the scattering invariant},
  author={Da Vela, Stefano and Begam, Nafisa and Dyachok, Danylo and Schaufele, Richard Santiago and Matsarskaia, Olga and Braun, Michal K and Girelli, Anita and Ragulskaya, Anastasia and Mariani, Alessandro and Zhang, Fajun and others},
  journal={The Journal of Physical Chemistry Letters},
  volume={11},
  number={17},
  pages={7273--7278},
  year={2020},
  publisher={ACS Publications}
}

@article{Schmidt1991,
author = {Schmidt, Paul W.},
doi = {10.1107/S0021889891003400},
issn = {00218898},
journal = {J. Appl. Crystallogr.},
month = {oct},
number = {pt 5},
pages = {414--435},
publisher = {International Union of Crystallography (IUCr)},
title = {{Small-angle scattering studies of disordered, porous and fractal systems}},
volume = {24},
year = {1991}
}

@article{horne2017,
  title={Rennet-induced coagulation of milk},
  author={Horne, David S and Lucey, John A},
  journal={Cheese},
  pages={115--143},
  year={2017},
  publisher={Elsevier}
}

@incollection{horne2020,
  title={Casein micelle structure and stability},
  author={Horne, David S},
  booktitle={Milk proteins},
  pages={213--250},
  year={2020},
  publisher={Elsevier}
}

@article{lucey2003,
  title={Invited review: Perspectives on the basis of the rheology and texture properties of cheese},
  author={Lucey, JA and Johnson, ME and Horne, DS},
  journal={Journal of dairy science},
  volume={86},
  number={9},
  pages={2725--2743},
  year={2003},
  publisher={Elsevier}
}

@incollection{fox2016,
  title={Enzymatic coagulation of milk},
  author={Fox, Patrick F and Guinee, Timothy P and Cogan, Timothy M and McSweeney, Paul LH},
  booktitle={Fundamentals of cheese science},
  pages={185--229},
  year={2016},
  publisher={Springer}
}

@article{bauland2024chapter,
  title={Enzymatic Gelation of Milk, Curd Draining and Cheese Yields},
  author={Bauland, Julien and Roustel, S{\'e}bastien and Faiveley, Marc and FAMELART, Marie-H{\'e}l{\`e}ne and Croguennec, Thomas and others},
  journal={Milk and Dairy Products: Some Challenges for the Dairy Industry},
  pages={129},
  year={2024},
  publisher={John Wiley \& Sons}
}

@article{salque2013,
  title={Earliest evidence for cheese making in the sixth millennium BC in northern Europe},
  author={Salque, M{\'e}lanie and Bogucki, Peter I and Pyzel, Joanna and Sobkowiak-Tabaka, Iwona and Grygiel, Ryszard and Szmyt, Marzena and Evershed, Richard P},
  journal={Nature},
  volume={493},
  number={7433},
  pages={522--525},
  year={2013},
  publisher={Nature Publishing Group UK London}
}

@article{holt1986,
  title={Effects of colloidal calcium phosphate content and free calcium ion concentration in the milk serum on the dissociation of bovine casein micelles},
  author={Holt, Carl and Davies, D Thomas and Law, Andrew JR},
  journal={Journal of Dairy Research},
  volume={53},
  number={4},
  pages={557--572},
  year={1986},
  publisher={Cambridge University Press}
}

@article{holt2013,
  title={Invited review: Caseins and the casein micelle: Their biological functions, structures, and behavior in foods},
  author={Holt, Carver and Carver, JA and Ecroyd, Heath and Thorn, DC},
  journal={Journal of dairy science},
  volume={96},
  number={10},
  pages={6127--6146},
  year={2013},
  publisher={Elsevier}
}

@article{bauland2024,
  title={Two-step aging dynamics in enzymatic milk gels},
  author={Bauland, Julien and Manna, Gouranga and Divoux, Thibaut and Gibaud, Thomas},
  journal={Physical review materials},
  volume={8},
  number={7},
  pages={L072601},
  year={2024},
  publisher={APS}
}

@article{narayanan2022,
  title={Performance of the time-resolved ultra-small-angle X-ray scattering beamline with the Extremely Brilliant Source},
  author={Narayanan, Theyencheri and Sztucki, Michael and Zinn, Thomas and Kieffer, J{\'e}r{\^o}me and Homs-Puron, Alejandro and Gorini, Jacques and Van Vaerenbergh, Pierre and Boesecke, Peter},
  journal={Applied Crystallography},
  volume={55},
  number={1},
  pages={98--111},
  year={2022},
  publisher={International Union of Crystallography}
}

@article{marchin2007effects,
  title={Effects of the environmental factors on the casein micelle structure studied by cryo transmission electron microscopy and small-angle x-ray scattering/ultrasmall-angle x-ray scattering},
  author={Marchin, St{\'e}phane and Putaux, Jean-Luc and Pignon, Fr{\'e}d{\'e}ric and L{\'e}onil, Jo{\"e}lle},
  journal={The Journal of chemical physics},
  volume={126},
  number={4},
  year={2007},
  publisher={AIP Publishing}
}

@article{holt2003substructure,
  title={Substructure of bovine casein micelles by small-angle X-ray and neutron scattering},
  author={Holt, C and De Kruif, CG and Tuinier, R and Timmins, PA},
  journal={Colloids and Surfaces A: Physicochemical and Engineering Aspects},
  volume={213},
  number={2-3},
  pages={275--284},
  year={2003},
  publisher={Elsevier}
}

@article{bouchoux2010squeeze,
  title={How to squeeze a sponge: Casein micelles under osmotic stress, a SAXS study},
  author={Bouchoux, Antoine and G{\'e}san-Guiziou, Genevi{\`e}ve and P{\'e}rez, Javier and Cabane, Bernard},
  journal={Biophysical journal},
  volume={99},
  number={11},
  pages={3754--3762},
  year={2010},
  publisher={Elsevier}
}

@incollection{HORNE2020213,
    title = {Chapter 6 - Casein micelle structure and stability},
    editor = {Mike Boland and Harjinder Singh},
    booktitle = {Milk Proteins},
    publisher = {Academic Press},
    edition = {Third},
    pages = {213-250},
    year = {2020},
    isbn = {978-0-12-815251-5},
    author = {David S. Horne},
}

@article{holt2022quantitative,
  title={Quantitative multivalent binding model of the structure, size distribution and composition of the casein micelles of cow milk},
  author={Holt, Carl and Carver, John A},
  journal={International Dairy Journal},
  volume={126},
  pages={105292},
  year={2022},
  publisher={Elsevier}
}

@article{takagi2022temperature,
  title={Temperature dependence of the casein micelle structure in the range of 10--40° C: An in-situ SAXS study},
  author={Takagi, Hideaki and Nakano, Tomoki and Aoki, Takayoshi and Tanimoto, Morimasa},
  journal={Food Chemistry},
  volume={393},
  pages={133389},
  year={2022},
  publisher={Elsevier}
}

@article{shukla2009structure,
  title={Structure of casein micelles and their complexation with tannins},
  author={Shukla, Anuj and Narayanan, Theyencheri and Zanchi, Dra{\v{z}}en},
  journal={Soft Matter},
  volume={5},
  number={15},
  pages={2884--2888},
  year={2009},
  publisher={Royal Society of Chemistry}
}

@article{raynes2024structure,
  title={Structure of biomimetic casein micelles: Critical tests of the hydrophobic colloid and multivalent-binding models using recombinant deuterated and phosphorylated $\beta$-casein},
  author={Raynes, Jared K and Mata, Jitendra and Wilde, Karyn L and Carver, John A and Kelly, Sharon M and Holt, Carl},
  journal={Journal of Structural Biology: X},
  volume={9},
  pages={100096},
  year={2024},
  publisher={Elsevier}
}

@article{day2017probing,
  title={Probing the internal and external micelle structures of differently sized casein micelles from individual cows milk by dynamic light and small-angle X-ray scattering},
  author={Day, Li and Raynes, JK and Leis, Andrew and Liu, LH and Williams, RPW},
  journal={Food Hydrocolloids},
  volume={69},
  pages={150--163},
  year={2017},
  publisher={Elsevier}
}

@article{de2012casein,
  title={Casein micelles and their internal structure},
  author={De Kruif, Cornelis G and Huppertz, Thom and Urban, Volker S and Petukhov, Andrei V},
  journal={Advances in colloid and interface science},
  volume={171},
  pages={36--52},
  year={2012},
  publisher={Elsevier}
}

@article{ingham2016revisiting,
  title={Revisiting the interpretation of casein micelle SAXS data},
  author={Ingham, Bridget and Smialowska, Alice and Erlangga, Gad Damar and Matia-Merino, Lara and Kirby, NM and Wang, Cheng and Haverkamp, Richard G and Carr, AJ},
  journal={Soft Matter},
  volume={12},
  number={33},
  pages={6937--6953},
  year={2016},
  publisher={Royal Society of Chemistry}
}

@article{pedersen2022model,
  title={A model on an absolute scale for the small-angle X-ray scattering from bovine casein micelles},
  author={Pedersen, Jan Skov and M{\o}ller, Thea Lykkegaard and Raak, Norbert and Corredig, Milena},
  journal={Soft Matter},
  volume={18},
  number={45},
  pages={8613--8625},
  year={2022},
  publisher={Royal Society of Chemistry}
}

@article{stothart1982small,
  title={Small-angle neutron scattering study of bovine casein micelles and sub-micelles},
  author={Stothart, Philip H and Cebula, Deryck J},
  journal={Journal of molecular biology},
  volume={160},
  number={2},
  pages={391--395},
  year={1982},
  publisher={Elsevier}
}

@article{mata2011structure,
  title={Structure of casein micelles in milk protein concentrate powders via small angle X-ray scattering},
  author={Mata, Jitendra P and Udabage, Punsandani and Gilbert, Elliot P},
  journal={Soft Matter},
  volume={7},
  number={8},
  pages={3837--3843},
  year={2011},
  publisher={Royal Society of Chemistry}
}

@article{pignon2004structure,
  title={Structure and rheological behavior of casein micelle suspensions during ultrafiltration process},
  author={Pignon, Fr{\'e}d{\'e}ric and Belina, Gabor and Narayanan, Theyencheri and Paubel, Xavier and Magnin, Albert and G{\'e}san-Guiziou, Genevi{\`e}ve},
  journal={The Journal of chemical physics},
  volume={121},
  number={16},
  pages={8138--8146},
  year={2004},
  publisher={American Institute of Physics}
}

@article{de2014structure,
  title={The structure of casein micelles: A review of small-angle scattering data},
  author={De Kruif, CG},
  journal={Applied Crystallography},
  volume={47},
  number={5},
  pages={1479--1489},
  year={2014},
  publisher={International Union of Crystallography}
}

@article{sorensen2013characterisation,
  title={Characterisation of fractionated skim milk with small-angle X-ray scattering},
  author={S{\o}rensen, Hanne and Pedersen, Jan Skov and Mortensen, Kell and Ipsen, Richard},
  journal={International Dairy Journal},
  volume={33},
  number={1},
  pages={1--9},
  year={2013},
  publisher={Elsevier}
}

@article{peyronel2020using,
  title={Using the USAXS technique to reveal the fat globule and casein micelle structures of bovine dairy products},
  author={Peyronel, Fernanda and Marangoni, Alejandro G and Pink, David A},
  journal={Food Research International},
  volume={129},
  pages={108846},
  year={2020},
  publisher={Elsevier}
}

@Inbook{DeKruif2003,
author={De Kruif, C. G.
and Holt, C.},
editor={Fox, P. F.
and McSweeney, P. L. H.},
title={Casein Micelle Structure, Functions and Interaction},
bookTitle={Advanced Dairy Chemistry---1 Proteins: Part A / Part B},
year={2003},
publisher={Springer US},
address={Boston, MA},
pages={233--276},
}

@article{trejo2011cryo,
  title={Cryo-transmission electron tomography of native casein micelles from bovine milk},
  author={Trejo, R and Dokland, T and Jurat-Fuentes, J and Harte, F},
  journal={Journal of dairy science},
  volume={94},
  number={12},
  pages={5770--5775},
  year={2011},
  publisher={Elsevier}
}

@article{li2018rheological,
  title={Rheological and structural properties of coagulated milks reconstituted in D2O: Comparison between rennet and a tamarillo enzyme (tamarillin)},
  author={Li, Zhao and Yang, Zhi and Otter, Don and Rehm, Christine and Li, Na and Zhou, Peng and Hemar, Yacine},
  journal={Food Hydrocolloids},
  volume={79},
  pages={170--178},
  year={2018},
  publisher={Elsevier}
}

@article{smith2020casein,
  title={Casein micelles in milk as sticky spheres},
  author={Smith, Gregory N and Brok, Erik and Christiansen, Morten Vormsborg and Ahrn{\'e}, Lilia},
  journal={Soft Matter},
  volume={16},
  number={43},
  pages={9955--9963},
  year={2020},
  publisher={Royal Society of Chemistry}
}

@article{fox2008casein,
  title={The casein micelle: Historical aspects, current concepts and significance},
  author={Fox, PF and Brodkorb, A},
  journal={International dairy journal},
  volume={18},
  number={7},
  pages={677--684},
  year={2008},
  publisher={Elsevier}
}

@article{holt2004equilibrium,
  title={An equilibrium thermodynamic model of the sequestration of calcium phosphate by casein micelles and its application to the calculation of the partition of salts in milk},
  author={Holt, Carl},
  journal={European Biophysics Journal},
  volume={33},
  number={5},
  pages={421--434},
  year={2004},
  publisher={Springer}
}

@article{gaucheron2005minerals,
  title={The minerals of milk},
  author={Gaucheron, Fr{\'e}d{\'e}ric},
  journal={Reproduction Nutrition Development},
  volume={45},
  number={4},
  pages={473--483},
  year={2005},
  publisher={EDP Sciences}
}

@article{lazzaro2020tailoring,
  title={Tailoring the structure of casein micelles through a multifactorial approach to manipulate rennet coagulation properties},
  author={Lazzaro, Fanny and Bouchoux, Antoine and Raynes, Jared and Williams, Roderick and Ong, Lydia and Hanssen, Eric and Lechevalier, Val{\'e}rie and Pezennec, St{\'e}phane and Cho, Hyun-Jung and Logan, Amy and others},
  journal={Food Hydrocolloids},
  volume={101},
  pages={105414},
  year={2020},
  publisher={Elsevier}
}

@article{li2022situ,
  title={In situ SAXS study of non-fat milk model systems during heat treatment and acidification},
  author={Li, Ruifen and J{\ae}ger, Tanja Christine and Rovers, Tijs AM and Svensson, Birte and Ipsen, Richard and Kirkensgaard, Jacob JK and Hougaard, Anni Bygvr{\aa}},
  journal={Food Research International},
  volume={157},
  pages={111292},
  year={2022},
  publisher={Elsevier}
}

@article{takagi2024saxs,
  title={A SAXS and USAXS study of the influence of pH on the casein micelle structure},
  author={Takagi, Hideaki and Nakano, Tomoki and Aoki, Takayoshi and Tanimoto, Morimasa},
  journal={Food Chemistry},
  volume={443},
  pages={138606},
  year={2024},
  publisher={Elsevier}
}

@article{lauzin2019effect,
  title={Effect of pH adjustment on the composition and rennet-gelation properties of milk concentrates made from ultrafiltration and reverse osmosis},
  author={Lauzin, A and B{\'e}rub{\'e}, A and Britten, M and Pouliot, Y},
  journal={Journal of Dairy Science},
  volume={102},
  number={5},
  pages={3939--3946},
  year={2019},
  publisher={Elsevier}
}

@article{sandra2012effect,
  title={Effect of soluble calcium on the renneting properties of casein micelles as measured by rheology and diffusing wave spectroscopy},
  author={Sandra, S and Ho, M and Alexander, M and Corredig, M},
  journal={Journal of Dairy Science},
  volume={95},
  number={1},
  pages={75--82},
  year={2012},
  publisher={Elsevier}
}

@article{dalgleish1979proteolysis,
  title={Proteolysis and aggregation of casein micelles treated with immobilized or soluble chymosin},
  author={Dalgleish, Douglas G},
  journal={Journal of Dairy Research},
  volume={46},
  number={4},
  pages={653--661},
  year={1979},
  publisher={Cambridge University Press}
}

@article{nogueira2020multiscale,
  title={Multiscale quantitative characterization of demineralized casein micelles: How the partial excision of nano-clusters leads to the aggregation during rehydration},
  author={Nogueira, Marcio H and Ben-Harb, Salma and Schmutz, Marc and Doumert, Bertrand and Nasser, Sarah and Derensy, Antoine and Karoui, Romdhane and Delaplace, Guillaume and Peixoto, Paulo PS},
  journal={Food Hydrocolloids},
  volume={105},
  pages={105778},
  year={2020},
  publisher={Elsevier}
}

\end{document}